\documentclass[usenatbib]{mn2e}

\usepackage{amsmath}
\usepackage{natbib}
\usepackage{epsfig}
\usepackage{txfonts}

\newcommand{\Mpc}{{\rm Mpc}}

\newcommand{\Planck}{{\sc Planck}}

\newcommand{\id}{{\,\rm d}}

\newcommand{\beq}{\begin{equation}}   %

\newcommand{\eeq}{\end{equation}}   %

\newcommand{\beqa}{\begin{eqnarray}}   %

\newcommand{\eeqa}{\end{eqnarray}}   %

\newcommand{\beal}{\begin{align}}

\newcommand{\enal}{\end{align}}

\newcommand{\bspl}{\begin{split}}

\newcommand{\espl}{\end{split}}

\newcommand{\bsub}{\begin{subequations}}

\newcommand{\esub}{\end{subequations}}

\newcommand{\bmulti}{\begin{multline}}   %

\newcommand{\beqm}{\begin{mathletters}}   %

\newcommand{\eeqm}{\end{mathletters}}   %

\newcommand{\me}{m_{\rm e}}

\newcommand{\Ne}{N_{\rm e}}

\newcommand{\Te}{T_{\rm e}}

\newcommand{\Tg}{T_{\gamma}}

\newcommand{\sigT}{\sigma_{\rm T}}

\newcommand{\pot}[2]{#1 \times 10^{#2}}

\newcommand{\Yp}{Y_{\rm p}}

\usepackage{color}

\usepackage{hyperref}
\usepackage{grffile}
\usepackage{graphics}
\usepackage{booktabs}

\title[Constraints on PMF from magnetic heating]
{Improved CMB anisotropy constraints on primordial magnetic fields from the post-recombination ionization history}
\author[D. Paoletti, J. Chluba, F. Finelli, J.~A.~Rubi\~{n}o-Mart\'{\i}n]
{D. Paoletti$^{1,2}$\thanks{E-mail: paoletti@iasfbo.inaf.it},~J. Chluba$^{3}$\thanks{E-mail: jens.chluba@manchester.ac.uk},
~ F. Finelli$^{1,2}$\thanks{E-mail: finelli@iasfbo.inaf.it} and 
J.~A.~Rubi\~{n}o-Mart\'{\i}n $^{4,5}$\thanks{E-mail: jalberto@iac.es} 
  \\
$^{1}$ INAF/OAS Bologna, Osservatorio di Astrofisica e Scienza dello Spazio, Area della ricerca CNR-INAF, via Gobetti 101, I-40129 Bologna, Italy\\
$^{2}$ INFN, Sezione di Bologna, Via Irnerio 46, I-40126, Bologna, Italy\\
$^{3}$ Jodrell Bank Centre for Astrophysics, Alan Turing Building, University of Manchester,Oxford Road Manchester M13 9PL, UK\\
$^{4}$ Instituto de Astrof\'{\i}sica de Canarias, C/V\'{\i}a L\'{a}ctea s/n, La Laguna, Tenerife, Spain\\
$^{5}$ Dpto. Astrof\'{i}sica, Universidad de La Laguna (ULL), E-38206 La Laguna, Tenerife, Spain
}

\date{{\vspace{-10.0mm} 2018}}

\begin{document}

\maketitle

\begin{abstract}
We investigate the impact of a stochastic background of Primordial Magnetic Fields (PMF) generated before recombination on the ionization history of the Universe and on the Cosmic Microwave Background radiation (CMB). Pre-recombination PMFs are dissipated during recombination and reionization via decaying MHD turbulence and ambipolar diffusion. This modifies the local matter and electron temperatures and thus affects the ionization history and Thomson visibility function. We use this effect to constrain PMFs described by a spectrum of power-law type extending our previous study (based on a scale-invariant spectrum) to arbitrary spectral index, assuming that the fields are already present at the onset of recombination. We improve previous analyses by solving several numerical issues which appeared for positively tilted PMFs indices.
We derive upper bounds on the integrated amplitude of PMFs due to the separate effect of ambipolar diffusion and MHD decaying turbulence and their combination. We show that ambipolar diffusion is relevant for $n_{\rm B}>0$  whereas for $n_{\rm B}<0$ MHD turbulence is more important. The bound marginalized over the spectral index on the integrated amplitude of PMFs with a sharp cut-off is $\sqrt{\langle  B^2 \rangle}<0.83$ nG. We discuss the quantitative relevance of the assumptions on the damping mechanism and the comparison with previous bounds. 
\end{abstract}



\section{Introduction}
\label{sec:intro}

Primordial magnetic fields (PMFs) generated prior to cosmological recombination provide an interesting window on the physics of the Early Universe and could have seeded 
the astrophysical large scale magnetic fields we observe in clusters and voids.
These PMFs leave imprints on the Cosmic Microwave Background (CMB)
through different mechanisms. PMF gravitate at the level of cosmological perturbations and source magnetically-induced perturbations. The comparison 
of theoretical predictions with different combinations of CMB data has been presented in several works 
\citep{Paoletti:2010rx,Shaw:2010ea,Paoletti:2012bb,Planck2013params,Ade:2015cva,Zucca:2016iur}, 
leading to constraints on the amplitude of PMFs smoothed at 1 Mpc of the order of few nG. The B-mode polarization induced by PMFs is also of great
interest for future CMB experiments \citep{2018arXiv180303230R,2018CQGra..35l4004P} 
PMFs also induce a Faraday rotation of CMB polarization, mixing E- and B-modes with an angle inversely proportional to the square of the frequency 
\citep{1996ApJ...469....1K,2009PhRvD..80b3009K,2011PhRvD..84d3530P}. 
At present, Faraday rotation leads to constraints which are weaker than those obtained by considering the gravitational effect, 
but represents a target for the future low-frequency polarization experiments and will help in disentangling the effects of helical and non-helical PMFs \citep{2009PhRvD..80b3009K,2011PhRvD..84d3530P,Ade:2015cva}.

 Together with the gravitational effect and the Faraday rotation of CMB polarization anisotropies, the presence of PMFs in the cosmological plasma prior to recombination may affect the thermal and ionization history of the Universe, significantly modifying the evolution of the cosmological plasma and consequently affecting both the CMB anisotropies and thermal spectrum.The dissipation of the PMFs by means of different mechanisms injects energy in the cosmological plasma heating it. The first direct consequence of this energy injection is the generation of distortions of the CMB absolute spectrum \citep{2000PhRvL..85..700J,Kunze2014, Wagstaff:2015jaa} \footnote{Note that the dissipation-induced distortions differ from those induced by cyclotron-radiation discussed in \cite{Burigana:2006ic}}. Both the distortions given by the dissipation of Alfven and magnetosonic waves and those generated from late (post-recombination) dissipation caused by MHD decaying turbulence and ambipolar diffusion are well below the COBE-FIRAS sensitivity \citep{Fixsen:1996nj}. Although current constraints on PMF from spectral distortions are not competitive with those from CMB anisotropies, future spectrometers like {\it PIXIE} \citep{Kogut:2011xw, Kogut2016SPIE} might represent an interesting avenue for improving the COBE-FIRAS limits.
 
 The presence of PMFs modifies the conditions of the pre-recombination plasma. In particular, on very small scales MHD turbulence may develop and then transfer energy between different scales \citep{Durrer:2013pga}. The MHD turbulence is one of the main ingredients in the evolution of the PMFs  when considering also the possible back reaction of the coupling with the fluid kinetic component; the presence of turbulence together with a time evolution of the magnetic energy density, can lead to a change in the spectrum of the PMFs \citep{Kahniashvili:2012vt,Saveliev:2012ea,Wagstaff:2013yna,Brandenburg:2014mwa,Wagstaff:2014fla,Brandenburg:2017neh,Reppin:2017uud, Trivedi2018}. If the PMFs are generated with an helical component simulations seem to indicate that the fields quite rapidly reach the maximal helical condition and that the evolution of the fields in presence of an helical component is modified \citep{Christensson:2000sp,Christensson:2002xu,Saveliev:2013uva,Kahniashvili:2016bkp,Brandenburg:2016odr}. 
 A full account of the MHD turbulence through the early Universe requires numerical simulations which up to this date are optimized for very small scales. CMB anisotropies on the other side are on very large scales (consider as an example comoving wavenumbers of the order of $k\sim 10^{-5}-0.1 \, {\rm Mpc^{-1}}$ leading to a problem of the matching between the different scales involved. In addition, the analysis with CMB data requires the predictions of the CMB anisotropies angular power spectra to be fed to the Markov Chain Monte Carlo pipeline. Thus, such massive predictions with Einstein-Boltzmann codes are not possible with current simulations set ups.
   
A full treatment which involves realistic simulations, CMB predictions and CMB data is still missing. It is therefore crucial to first assess the importance of the effect of PMFs on the thermal and ionization history of the Universe especially in the light of the recent blossoming of CMB data. Recent works \citep{Kunze2014, 2015JCAP...06..027K, 2015MNRAS.451.2244C, Ade:2015cva} have considered the post-recombination dissipative effects and derived an upper 
limit on the PMFs integrated amplitude for a nearly scale-invariant and negative indices \citep{2015JCAP...06..027K}  stochastic background at the nG level, tighter than those derived on the basis of the gravitational effects only. 
These analyses do not involve full MHD simulations but use analytical energy injections rates \citep{Sethi2005, 2005PhRvD..72b3004S,Sethi2009} which are included into the Einstein-Boltzmann codes to derive the CMB anisotropies angular power spectra. As it is usually done for the gravitational effect the ideal MHD limit is assumed where the PMFs are frozen in the plasma and we neglect possible back-reaction of the fluid onto the fields considering these effects as second order. 
However, as stressed previously \citep{2015MNRAS.451.2244C, Ade:2015cva}, significant uncertainties exist in the description of the heating rates and consequently the derived constraints.
The main scope of this paper is the improvement of previous analyses \citep{Kunze2014, 2015JCAP...06..027K, 2015MNRAS.451.2244C, Ade:2015cva} curing numerical aspects which prevented the study of blue tilted spectrum PMFs. Although incomplete, the approximate treatment presented here provides an important intermediate step towards a full ambitious analysis.

We derive the CMB constraints on a stochastic background of PMFs by their impact on the modified ionization history and anisotropies angular power spectra beyond the nearly-scale invariant case previously reported \citep[e.g.,][]{Ade:2015cva,2015MNRAS.451.2244C}. Constraints for PMF spectral indices $n_{\rm B}=-1.5$ and $-2.5$ were already obtained by \citet{2015JCAP...06..027K}. Here we extend the analysis to arbitrary spectral index and improve the treatment including subtle effects. 
We improved the numerical accuracy of the recombination code {\tt Recfast++} \citep{2011MNRAS.412..748C}, which includes the heating effect of PMFs by means of two different methods dedicated specifically to MHD turbulence and to ambipolar diffusion.
In order to maximize the numerical stability of {\tt CAMB}, following \citet{Hart2018prep}, we also enhanced the time-step settings during recombination which hampered the precision of the obtained CMB power spectra at larges scales, leading to a slower convergence of MCMC chains.  

The paper is organized as follows. In section 2 we describe the details of a stochastic background of PMFs and 
of the induced modified ionization history. In section 3 we describe the impact of the MHD decaying turbulence and of the ambipolar diffusion on the CMB power spectra. We present the constraints from \Planck\ 2015 data in section 4. In section 5 we discuss our results and we draw our conclusions in section 6. 
In appendix A we describe the implications of our results on the commonly adopted amplitude of PMF smoothed at 1 Mpc scale.

\section{Impact of primordial magnetic fields on the post-recombination ionization history}
We consider a fully inhomogeneous stochastic background of non-helical PMFs which in Fourier space is described by:
\begin{equation}
\langle B_i({\bf k}) B_j^*({\bf k}')\rangle=(2\pi)^3 \delta({\bf k}-{\bf k}')
(\delta_{ij}-\hat k_i\hat k_j) \frac{P_B(k)}{2}
\end{equation}
where the magnetic power spectrum is \footnote{$n_{\rm B} > -3$ to avoid infrared divergences} $P_B(k)=A_{\mathrm B} k^{n_{\rm B}}$.Since we are interested in the relevant scales for CMB anisotropies we consider the ideal MHD limit in which the PMF energy density behaves as a relativistic component  $\rho_{\mathrm{B}}({\bf{x}},\tau)=\frac{\rho_{\mathrm{B}}({\bf{x}})}{a^4(\tau)}$ with $B({\bf{x}},\tau)=\frac{B({\bf{x}})}{a^2(\tau)}$. We neglect higher order non-linear effects of the interaction of the magnetic field with the fluid which may lead to a different evolution of the energy density of the fields on small scales, see for example \cite{Saveliev:2012ea, Saveliev:2013uva,Brandenburg:2016odr}.

Radiation viscosity damps PMFs at a damping scale $k_{\mathrm D}$ \citep{Jedamzik1998, Subramanian1998}:
\begin{equation}
{\bf \frac{\bf k_{\rm D}}{\bf \Mpc^{-1}}} = 
\frac{\sqrt{5.5\times 10^4} (2\pi)^\frac{n_{\rm B}+3}{2}}{\sqrt{\langle  B^2 \rangle} /nG \sqrt{\Gamma[(n_{\rm B}+5)/2]}}\, \sqrt{h \,\frac{\Omega_b h^2}{0.022}}\,.
\label{KD}
\end{equation}
In this paper, we choose to model this damping by imposing a sharp cut-off at the scale $k_{\mathrm D}$ to regularize ultraviolet divergencies in integrated quantities, 
as done in the study of the PMFs gravitational effects.
We therefore define the root mean square as:
\begin{equation}
\langle B^2 \rangle = \frac{A_{\mathrm B}}{2 \pi^2} \int_0^{k_{\mathrm D}} dk k^{2+n_{\mathrm B}} = \frac{A_B}{2 \pi^2 (n_{\rm B}+3)} k_{\rm D}^{n_{\rm B}+3}\,.
\label{rms_sharp}
\end{equation}
Note that in our previous paper \citep{2015MNRAS.451.2244C} we considered a Gaussian smoothing as in \cite{2015JCAP...06..027K} to regularize the integrated amplitude of the stochastic background. According to \citet{Sethi2005}, the heating due to PMFs to the electron temperature equation is modelled as:
\beal
\label{eq:dT_dt}
\frac{\id \Te}{\id t}=- 2 H \Te 
+ \frac{8 \sigT\Ne \,\rho_\gamma}{3\me c N_{\rm tot}} (\Tg-\Te) + 
\frac{\Gamma}{(3/2)k N_{\rm tot}} \,,
\end{align}
where $H(z)$ denotes Hubble rate, $N_{\rm tot} =N_{\rm H}(1+f_{\rm He}+X_{\rm e})$ 
the number density of all ordinary matter particles that share the thermal energy, 
beginning tightly coupled by Coulomb interactions; $N_{\rm H}$ is the number density 
of hydrogen nuclei, $f_{\rm He}\approx \Yp/ 4(1-\Yp)\approx 0.079$ for helium mass 
fraction $\Yp=0.24$; $X_{\rm e}=N_{\rm e}/N_{\rm H}$ denotes the free electron 
fraction and $\rho_\gamma=a_{\rm R} \Tg^4\approx 0.26 \, {\rm eV} (1+z)^4$ the CMB energy density. 
The first term in Eq.~\eqref{eq:dT_dt} describes the adiabatic cooling of matter due to the Hubble expansion, 
while the second term is caused by Compton cooling and heating. 
The last term accounts for the PMF heating due to the sum of the decaying magnetic turbulence ($\Gamma_{\rm turb}$) and ambipolar ($\Gamma_{\rm amb}$), respectively.

We review in the following the approach of the aforementioned heating terms and describe the regularization and numerical improvements we provide with respect to previous treatments \citep{2015JCAP...06..027K, 2015MNRAS.451.2244C}.

\begin{figure}
\centering
\includegraphics[width=\columnwidth]{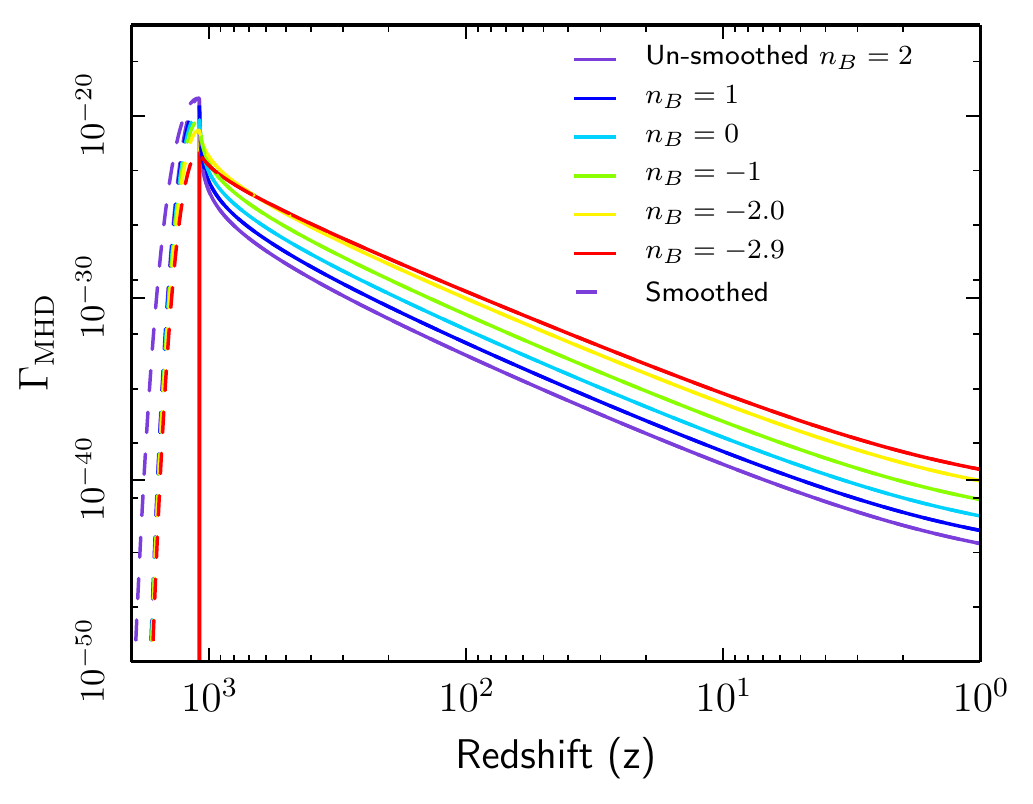}
\caption{Comparison of the MHD decaying turbulence heating rate for unsmoothed (solid lines) 
and smoothed rates (dashed lines), the different colors stand for the 
spectral indices as in the legend. The PMFs amplitude is set to  $\sqrt{\langle  B^2 \rangle}=0.4$ nG.}
\label{fig:MHDRate}
\end{figure}

\begin{figure}
\centering
\includegraphics[width=\columnwidth]{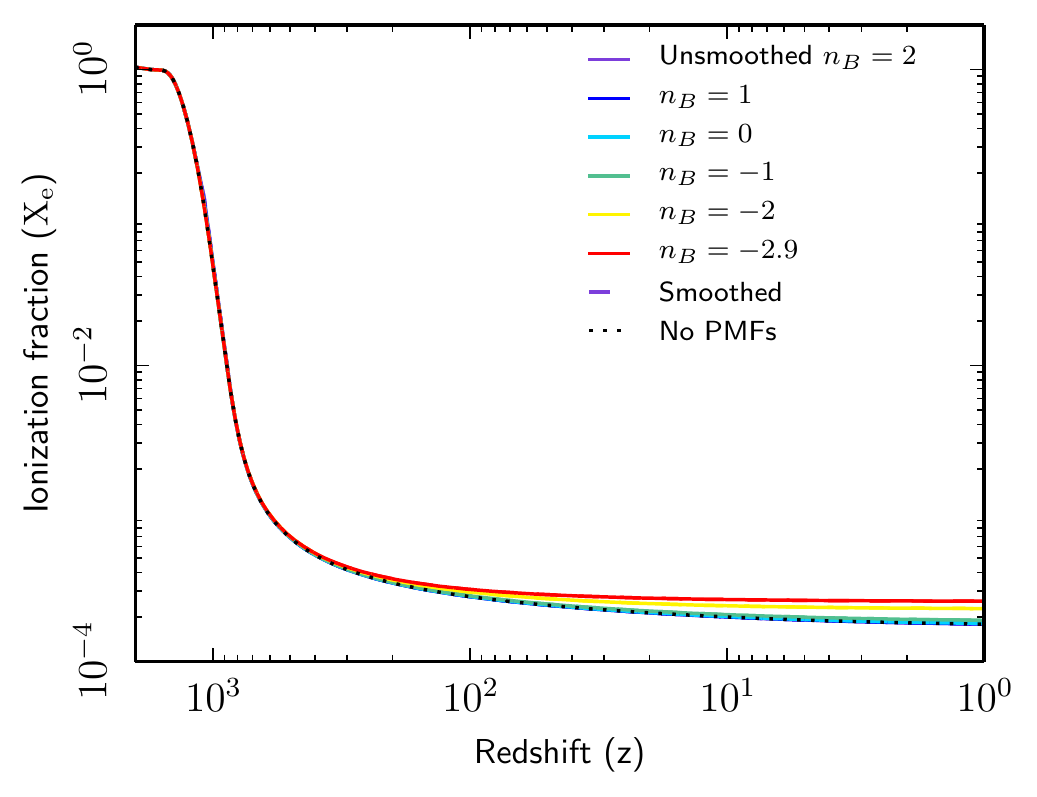}
\caption{Comparison between the smoothed and unsmoothed rate impacts on the ionization fraction. The PMF amplitude is set to 
$\sqrt{\langle  B^2 \rangle} = 0.4$ nG. Colors represent the different spectral indices as in the legend.}
\label{fig:xe}
\end{figure}

\subsection{Decaying MHD turbulence}
On scales smaller than the magnetic Jeans scale, PMFs may be subject to non-linear effects and develop MHD turbulence. 
Before recombination the radiation viscosity over-damps the velocity fluctuations maintaining 
the Reynold number small. After recombination, the sudden drop of radiation viscosity allows 
for the development of large Reynold number and for the transfer of energy from large towards smaller scales, dissipating energy. 
The dissipation of the fields injects energy into the plasma, with a rate that can be approximated as \citep{Sethi2009}:
\beal
\Gamma_{\rm turb}=\frac{3 m}{2}\, 
\frac{\left[\ln \left(1+\frac{t_i}{t_{\rm d}}\right)\right]^m}{\left[\ln \left(1+\frac{t_i}{t_{\rm d}}\right)
+ \frac{3}{2} \ln \left( \frac{1+z_i}{1+z}\right)\right]^{m+1}} H(z)\,\rho_{\rm B}(z),
\label{Eqn:rate}
\end{align}
with the parameters $m=2(n_{\rm B}+3)/(n_{\rm B}+5)$, 
$t_i/t_{\rm d}\approx 14.8 (\langle  B^2 \rangle^{1/2} / {\rm nG})^{-1}(k_{\rm D}/ \Mpc^{-1})^{-1}$, 
and magnetic field energy density  $\rho_{\rm B}(z)=\langle  B^2 \rangle (1+z)^4/ (8\pi) 
\approx \pot{9.5}{-8} (\langle  B^2 \rangle/ {\rm nG}^{2})\,\rho_\gamma(z)$.

\subsubsection{Regularizing around recombination}
\label{sec:reg_func}
Following previous approaches  \citep{Ade:2015cva,2015MNRAS.451.2244C} the heating term due to decaying magnetic turbulence in Eq. (\ref{Eqn:rate}) switches on abruptly at $z_i\sim1088$. 
Although the rate is a continuous function, the cusp at $z=1088$,  shown in Fig.\ref{fig:MHDRate},
creates numerical issues for the derivatives within the modified recombination code we have developed to include PMFs. 
The decaying magnetic turbulent rate in Eq. (\ref{Eqn:rate}) is weakly coupled to the time evolution of the electron temperature 
in Eq. (\ref{eq:dT_dt}) for $n_\mathrm{B} \approx -3$ and therefore in this case the abrupt switch on is numerically tolerable. 
This is the reason why previous studies in \cite{Ade:2015cva,2015MNRAS.451.2244C} were restricted to $n_\mathrm{B} = -2.9$.
In order to extend our study to different spectral indices, we introduce a smoothing of the decaying magnetic turbulent rate 
which includes a Gaussian suppression before recombination. 
In particular we consider the phenomenological model:
\begin{itemize}
\item for $z<z_i\sim1088$, Eq.~\eqref{Eqn:rate};
\item for $z_i \le z \le 1.001 z_i$ polynomial to smooth the derivative at $z_i$ and make it zero at $1.001 z_i$;
\item for $z>1.001 z_i$ Gaussian suppression to model the onset of turbulent heating.
\end{itemize}
More recent 3D simulations suggest a slow power-law behavior for the onset of turbulent heating \citep{Trivedi2018}, however, here we remain as closely as possible within the old framework, leaving a study of these improved magnetic heating rate calculations to future work. The smoothed rate is shown in Fig.\ref{fig:MHDRate} together with the unsmoothed one for different spectral indices and fixed $\sqrt{\langle  B^2 \rangle}=0.4~{\rm nG}$.
Note how the regularization we have applied affects only the redshifts around recombinations, it does not affect later epochs.
In Fig.\ref{fig:xe} we show the effect of the smoothing on the ionization fraction. We note how the smoothing has a negligible impact on the ionization fraction, we will see how this is reflected in a negligible impact on the angular power spectra with the MHD heating decaying turbulence effect.

\begin{figure*}
\centering
\includegraphics[width=0.9\columnwidth]{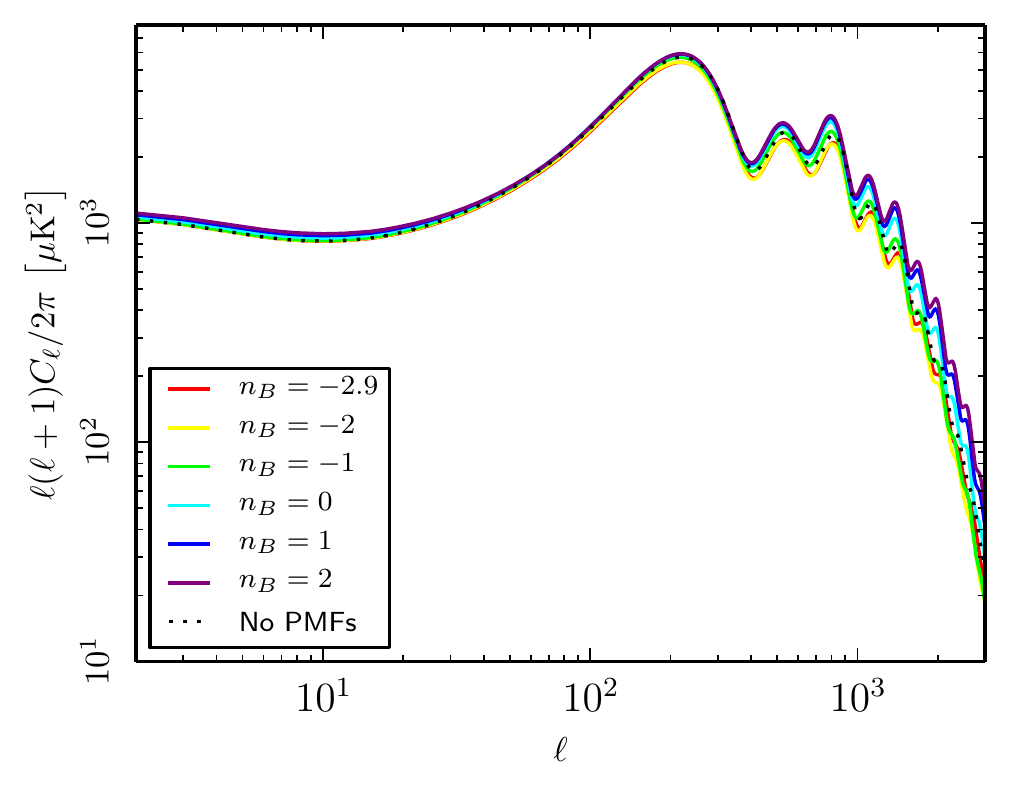}
\hspace{3mm} 
\includegraphics[width=0.9\columnwidth]{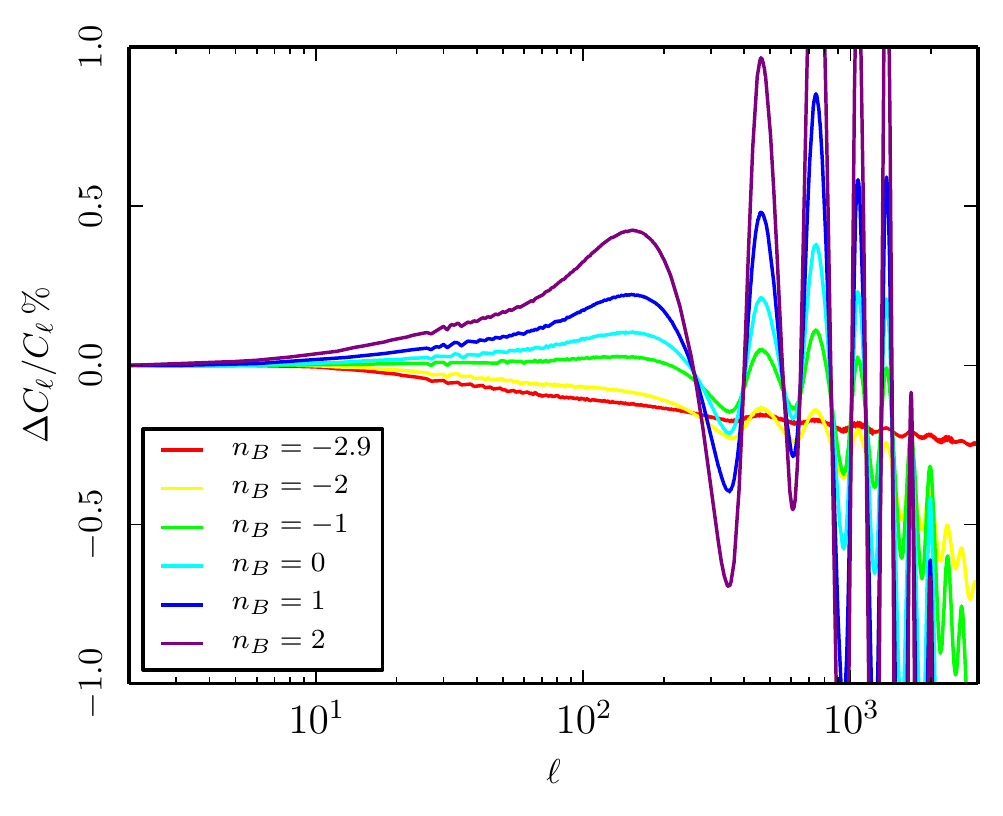} \\ 
\includegraphics[width=0.9\columnwidth]{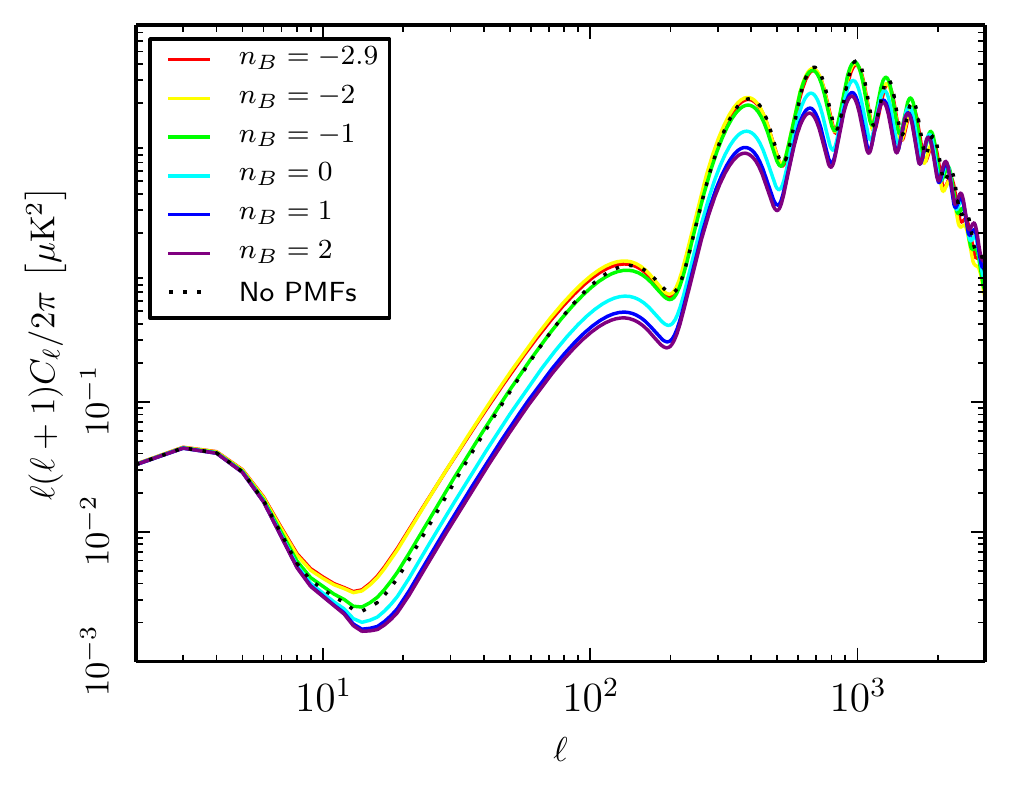}
\hspace{3mm} 
\includegraphics[width=0.9\columnwidth]{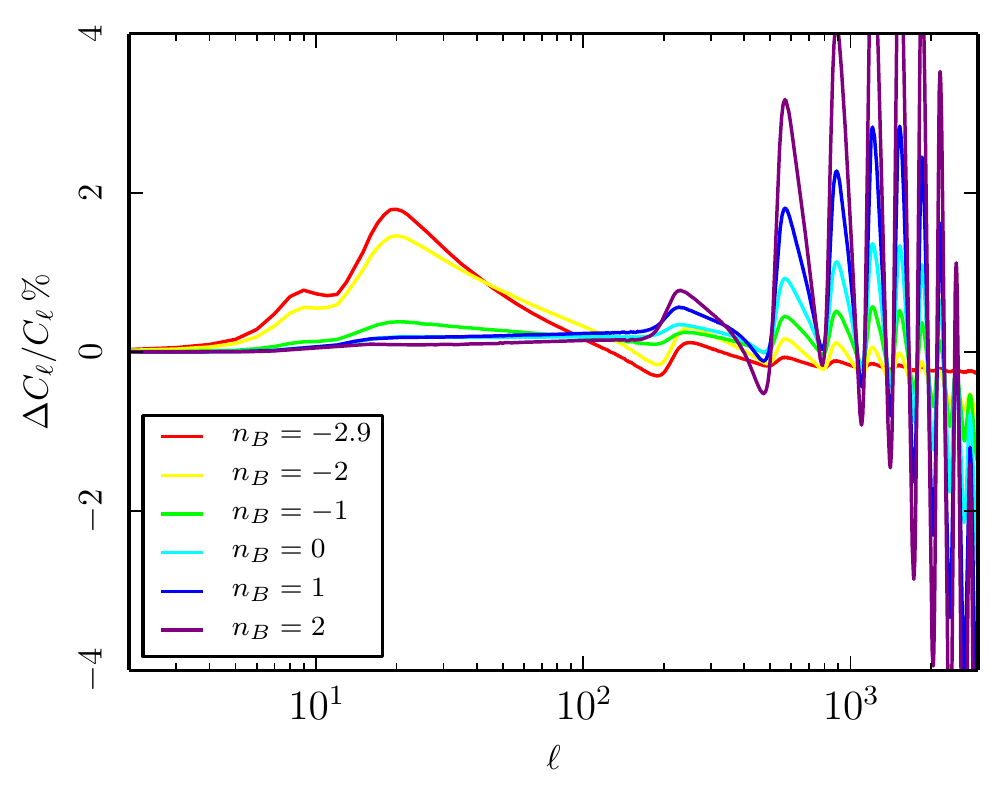} 
\caption{In the left column we present the angular power spectra with MHD decaying turbulence effect. To illustrate the effect we increased the amplitude of the field with respect to the right column to $\sqrt{\langle  B^2 \rangle}$=4 nG. In the right column we present the relative differences with and without MHD decaying turbulence effect, $\sqrt{\langle  B^2 \rangle}$=0.4 nG, of the CMB anisotropy angular power spectra in temperature and polarization.
}
\label{fig:ClN}
\label{fig:ClAs}
\end{figure*}

\begin{figure}
\centering
\includegraphics[width=0.9\columnwidth]{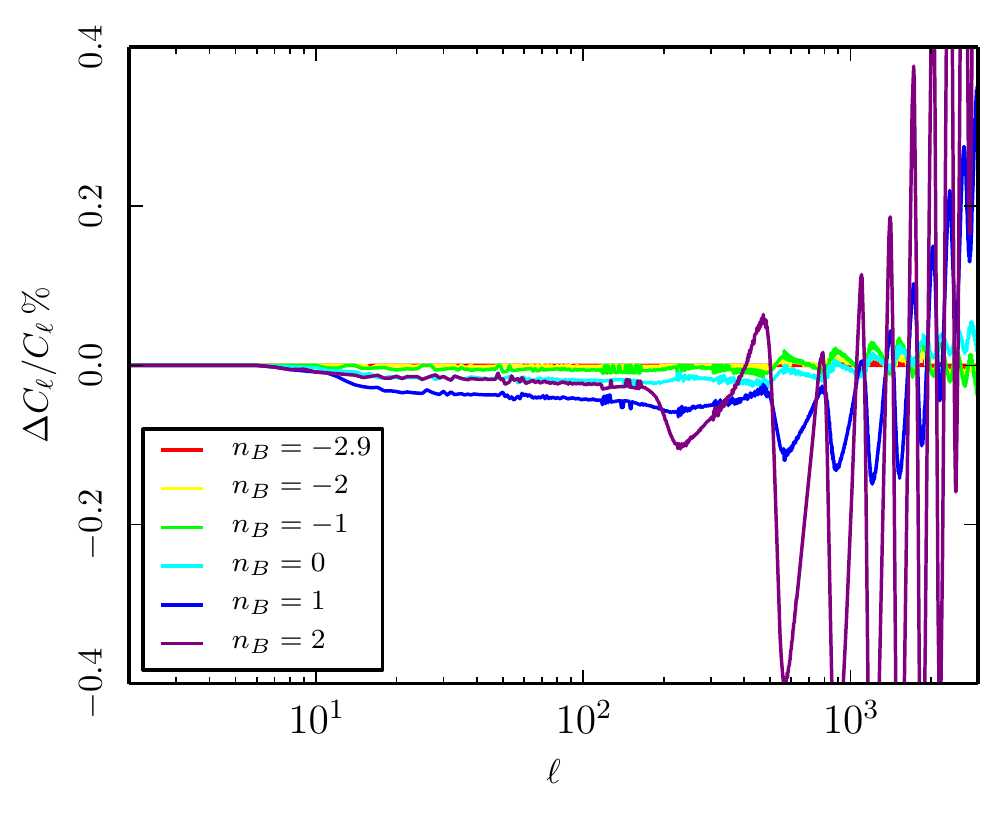}
\caption{Relative difference of smoothed and unsmoothed power spectra for temperature anisotropies, the colors are in the legend.}
\label{fig:ClSNS}
\end{figure}

\subsection{Ambipolar diffusion}
The ambipolar diffusion arises in partially ionized plasmas in the presence of magnetic fields. 
Being the cosmological plasma only partially ionized after recombination and since the Lorentz force induced by PMFs acts only on the ionized 
component, there is a difference  between the velocity of ions and that of neutral atoms. 
Collisions between the two dissipate this difference and rapidly thermalize the energy which is transferred 
to the neutral component. This mechanism dissipates the PMFs and heats the plasma,  
if the heating is strong this effect may also change the ionization fraction evolution itself.
To capture the effect of heating by ambipolar diffusion we use the approximation \citep{Sethi2005, Schleicher2008b}:
\beal
\Gamma_{\rm am}\approx \frac{(1-X_{\rm p})}{\gamma X_{\rm p}\, \rho_{\rm b}^2} \left< {\bf L}^2\right>
\end{align}
where $\left< {\bf L}^2\right>=|(\nabla \times {\bf B})\times {\bf B}|^2/(4\pi)^2$ denotes
the average square of the Lorentz-force 
$\rho_{\rm b}=m_{\rm H} N_{\rm b}$ the baryon {\it mass} density with baryon number density $N_{\rm b}$.
and $X_{\rm p}=N_{\rm p}/N_{\rm H}$ the coupling between the ionized and neutral component. The coupling coefficient
is given by $\gamma=\left<\sigma \varv\right>_{H\,H^+}/2m_{\rm H}$ with
$\left<\sigma \varv\right>_{H\,H^+}\approx \pot{6.49}{-10} (T/{\rm K})^{0.375} {\rm cm}^3 \, {\rm s}^{-1}$.
For $-2.9 < n_{\rm B} < 2$, the integral for the Lorentz force according to a sharp cut-off prescription is:
\beal
\label{eq:L_def}
|(\nabla\times B)\times B|^2 &= 16\pi^2 \rho^2_{\rm B}(z)\,l_{\rm D}^{-2}(z)\,g_L(n_{\rm B}+3) \\
%
\label{eq:g_L_K}
g_L (x) &= 0.6615 [1 - 0.1367 x + 0.007574 x^2]\, x^{0.8874} \,.
\end{align}
with $l_{\rm D}=a/k_{\rm D}$. Note that the Lorentz force is computed in this paper for a sharp cut-off, consistently with the rms amplitude 
of the stochastic background in Eq. (\ref{rms_sharp}), whereas 
in our previous paper \citep{2015MNRAS.451.2244C} we instead adopted a Gaussian smoothing to compare with the results in \cite{Kunze2014}. 

In order to solve the numerical issues with the ambipolar diffusion effect for PMFs with positive spectral indices we also 
improved the numerical integration of {\tt Recfast++} \citep{2011MNRAS.412..748C}, adding an explicit solve of the linear algebra problem appearing at each time-step in the ordinary differential equation problem. This improved the numerical stability at the onset of ambipolar diffusion around redshift $z\simeq 100-200$

\begin{figure*}
\centering
\includegraphics[width=0.9\columnwidth]{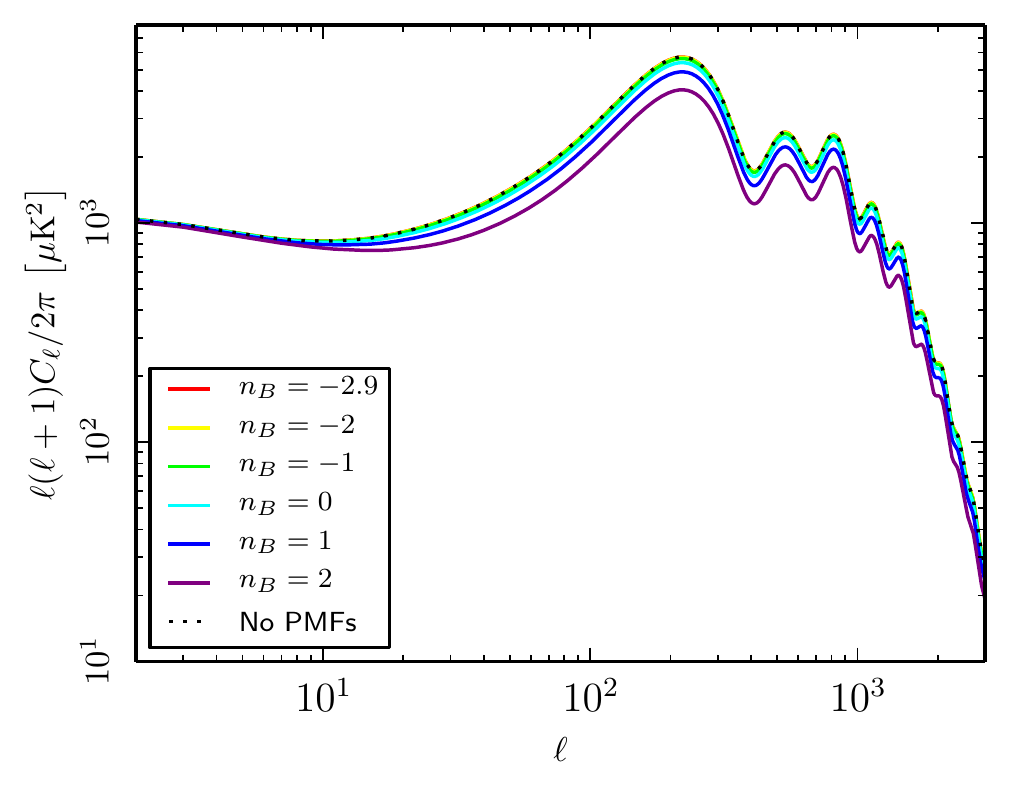}
\hspace{3mm}
\includegraphics[width=0.9\columnwidth]{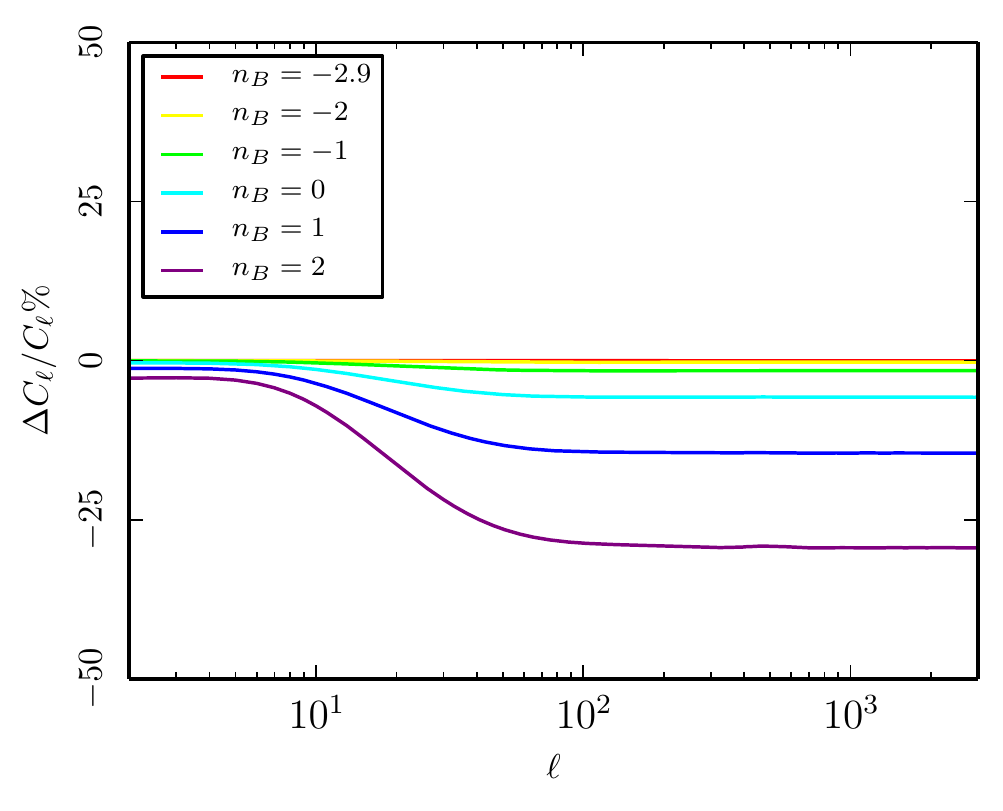}\\ 
\includegraphics[width=0.9\columnwidth]{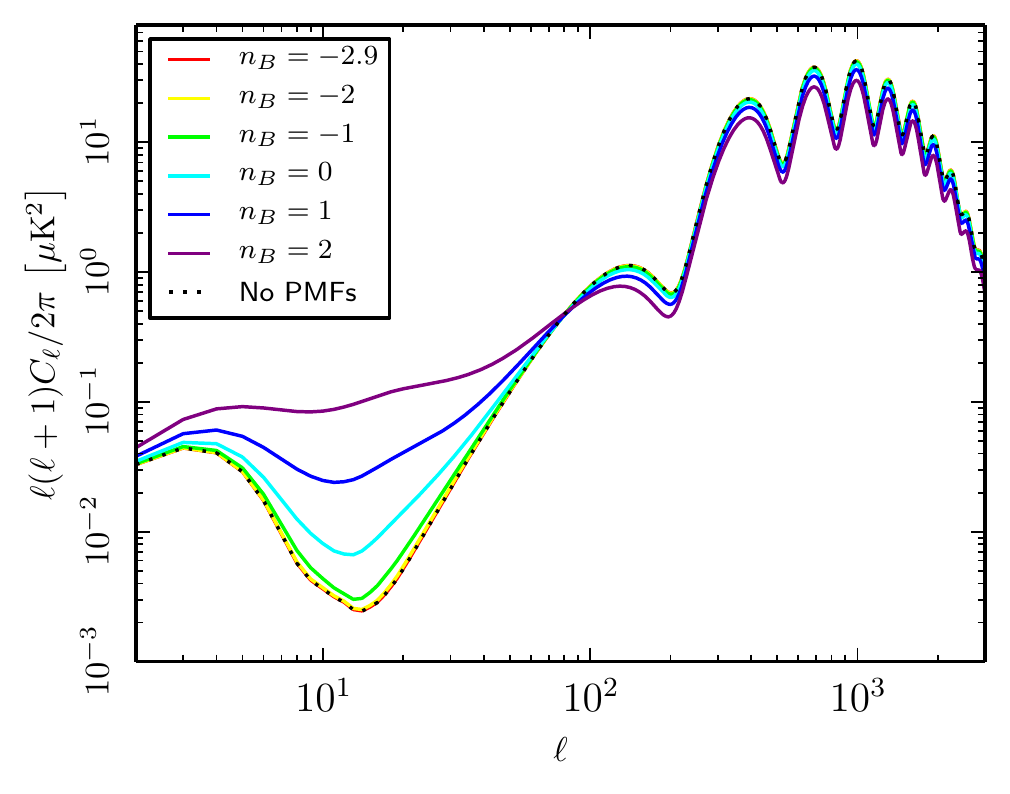}
\hspace{3mm}
\includegraphics[width=0.9\columnwidth]{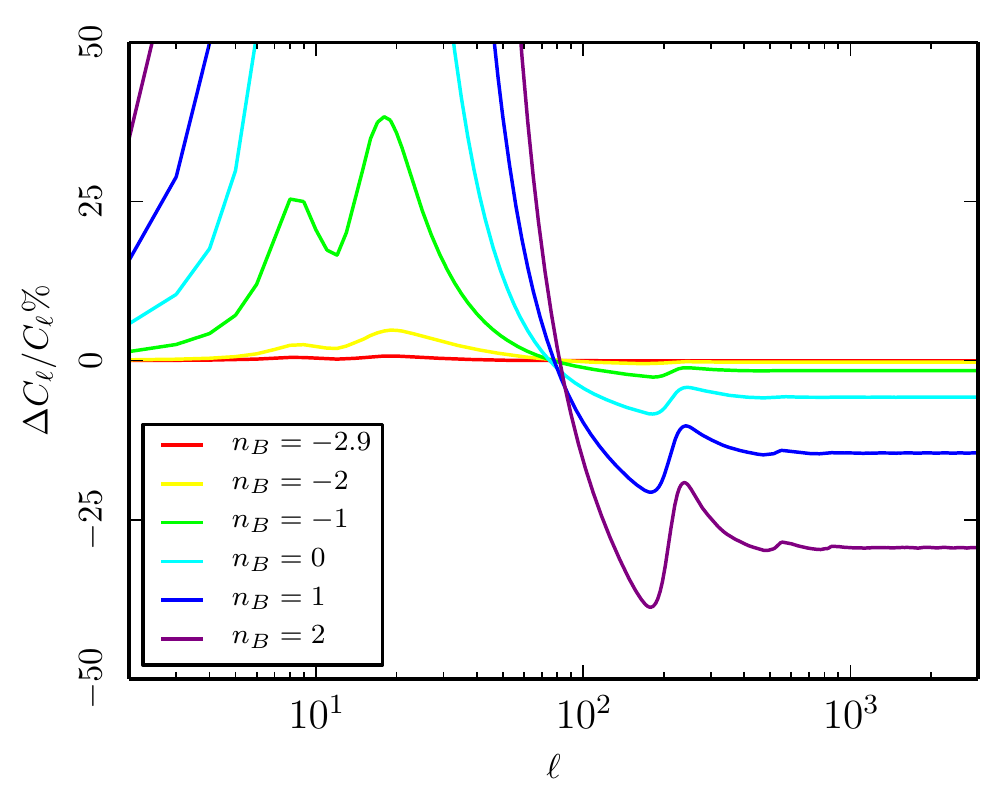}
\caption{On the left column we show the impact of ambipolar diffusion on the CMB angular power spectra for PMFs with an amplitude  $\sqrt{\langle  B^2 \rangle}$=0.4 nG for different spectral indices compared with case without PMF heating in black. The upper panel is $TT$ the lower panel is $EE$. Colors represent the different spectral indices. On the right column instead we show the relative difference of the case with and without the ambipolar diffusion for PMFs with amplitude of  $\sqrt{\langle  B^2 \rangle}$=0.4 nG for different spectral indices. The upper panel is $TT$ the lower panel is $EE$. Colors represent the different spectral indices.}
\label{fig:ClA}
\end{figure*}

\section{CMB angular power spectra}
We now briefly present the impact of the PMF dissipation on the CMB angular power spectra in temperature and polarization. These are very similar to previous computations; however, the numerical noise which was present at large angular scales is eliminated thanks to the improved time-sampling inside {\tt CAMB}.
  
\subsection{The impact of MHD decaying turbulence}
We start by describing the MHD decaying turbulence effect. In the left column of  Fig.~\ref{fig:ClAs} we illustrate the effect on the temperature and E-mode polarization angular power spectra for $\sqrt{\langle  B^2 \rangle}$=4 nG, note that for this specific figure we have increased the amplitude of the fields with respect to the others of this section in order to visually enhance the effect. In the right column of Fig.~\ref{fig:ClAs} we present the relative differences of the angular power spectra which include and 
do not include the MHD turbulence effect, note that for these figures the amplitude of the fields is $\sqrt{\langle  B^2 \rangle}$=0.4 nG, which is closer to the value obtained in the data analysis. We note in particular a strong effect on the E-mode polarization at intermediate and small angular scales and a sub-percent effect in temperature on small angular scales. In contrast to previous computations \citep[e.g.,][]{Kunze2014, Ade:2015cva}, the effect at large angular scales is less pronounced. This is because following \citet{Hart2018prep}, we significantly increased the time-sampling\footnote{This is controlled by the parameter {\tt dtaurec}.} in {\tt CAMB} ($\simeq 100$ times) to better resolve the onset of heating around $z\simeq 1088$. This  improvement eliminates the dependence of the angular power spectrum on large scales on the accuracy parameters making the Boltzmann code very stable as can be seen in Fig. \ref{fig:ClAs} where large scales do not show any feature.

We have described the regularization function we apply in order to solve numerical issues of the MHD turbulence treatment for positive spectral indices (Sect.~\ref{sec:reg_func}). 
In Fig.~\ref{fig:ClSNS} we show the relative differences of the cases with and withouth the smoothing for  $\sqrt{\langle  B^2 \rangle}=0.4$ nG. The effect of our regularization remains at the sub-percent level in all considered cases, with the largest effect seen for $n_{\rm B}=2$.
Please note that for $n_{\rm B}=2$ an amplitude $\sqrt{\langle  B^2 \rangle}=0.4$ for the root mean square of the PMFs is already ruled out by data. For indices smaller or equal zero the angular power spectra do not show any significant dependence on the chosen regularization scheme. We can therefore conclude that for the amplitudes we are able to constrain with this methodology the application of the regularization of the rate does not affect the results of the analysis.

\subsection{Ambipolar diffusion}
We now proceed by illustrating the effect of the ambipolar diffusion on the CMB angular power spectra. In the left column of  Fig.~\ref{fig:ClA} we show the angular power spectrum in temperature and E-mode polarization with the effect of ambipolar diffusion compared with the case without PMFs. We considered different spectral indices and PMFs with an amplitude of $\sqrt{\langle  B^2 \rangle}=0.4$ nG as in the previous case. For more clarity, in the right column of Fig.~\ref{fig:ClA} we show the relative difference between the ambipolar diffusion case and the case without PMF contribution. The main effect of ambipolar diffusion heating is a reduction of the overall amplitude of the $TT$ power spectra at intermediate and small scales ($\ell \gtrsim 10$). In contrast, for the $EE$ power spectra, the effect is more pronounced at large angular scales around the reionization bump which for very blue indices of the order of $n_{\rm B}=1-2$ is strongly suppressed (cf., Fig.~\ref{fig:ClA} ). This illustrates that the main effect of ambipolar diffusion heating is an increase of the total Thomson optical depth to last scattering. The overall features are consistent with previous studies \citep[e.g.,][]{Kunze2014}.

\begin{figure*}
\centering
\includegraphics[width=0.9\columnwidth]{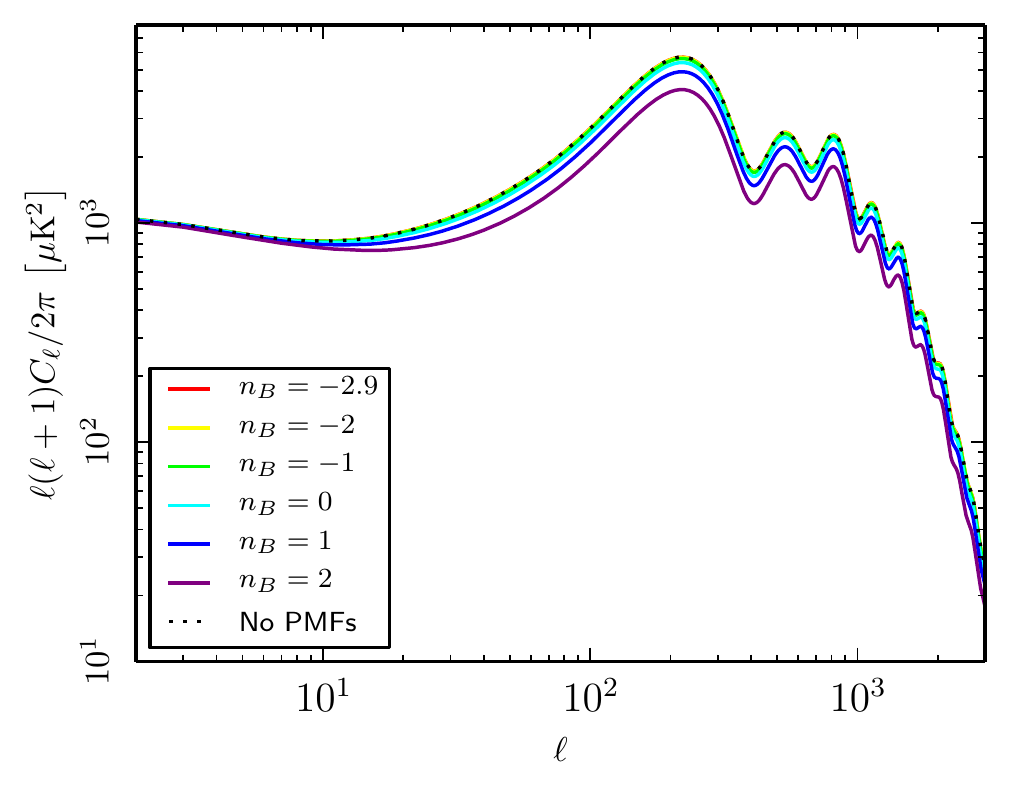}
\hspace{3mm}
\includegraphics[width=0.9\columnwidth]{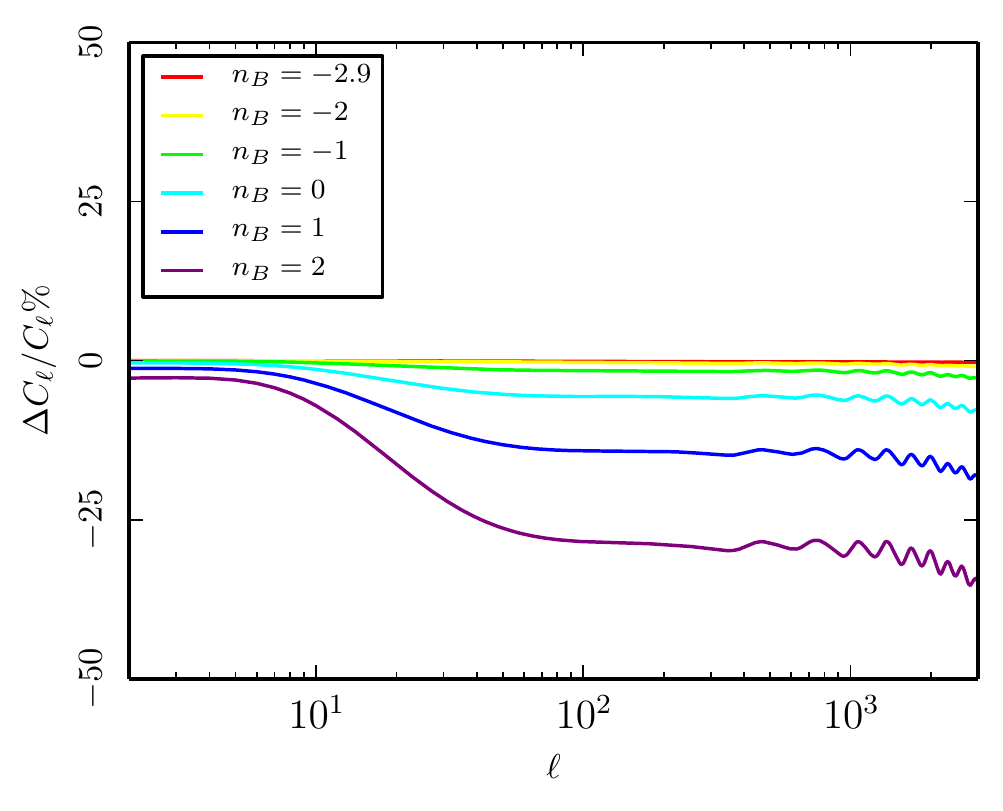}
\\  
\includegraphics[width=0.9\columnwidth]{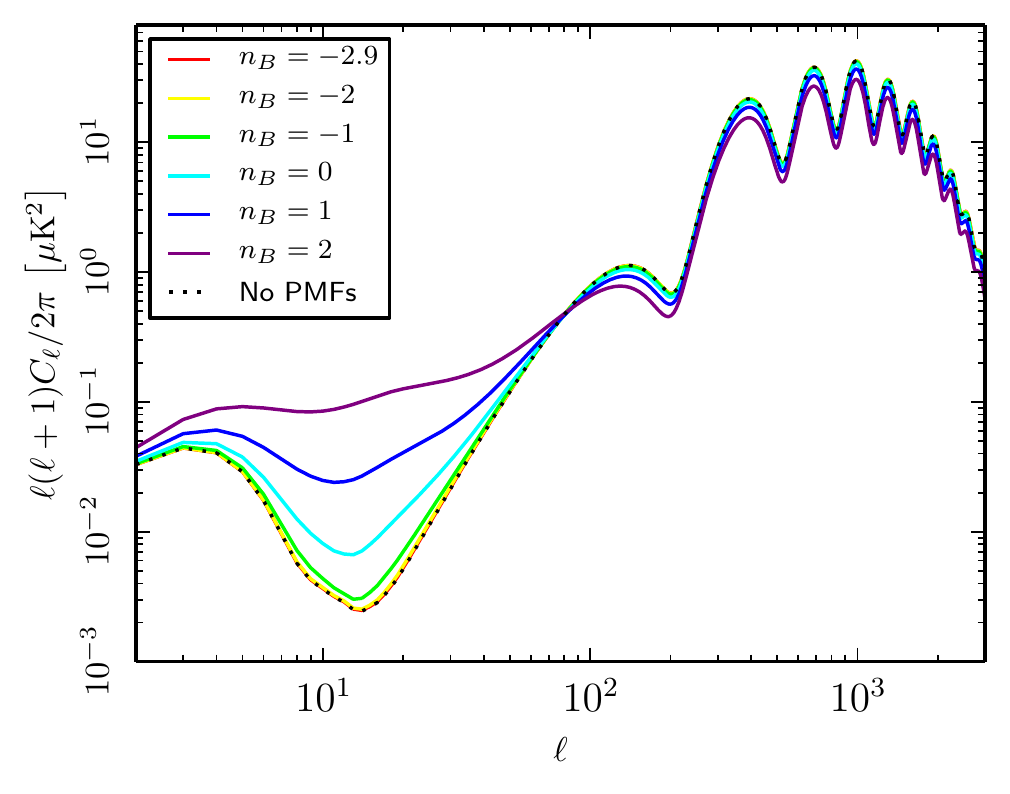}
\hspace{3mm}
\includegraphics[width=0.9\columnwidth]{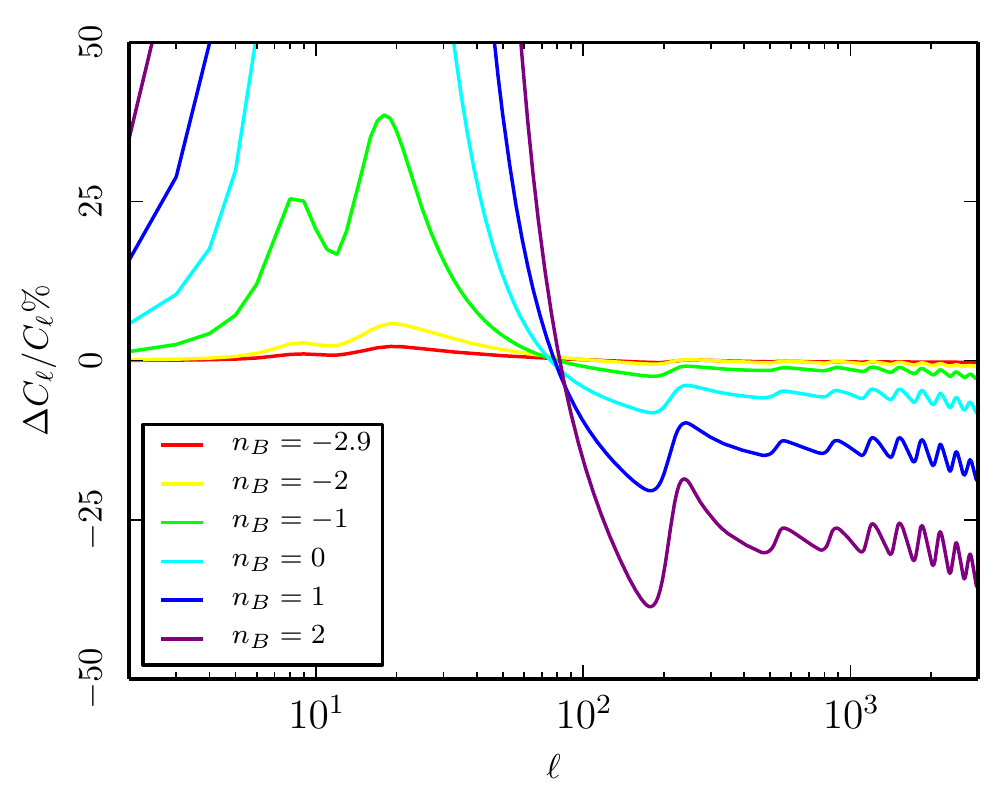}
\caption{In the left column we show the combined effect on the CMB angular power spectra of PMFs with an amplitude $\sqrt{\langle  B^2 \rangle}$=0.4 nG for different spectral indices compared with case without PMFs (in black). The upper panel is $TT$ the lower panel is $EE$. Colors represent the different spectral indices. In the right column we present the relative difference of the angular power spectra of the cases with and without heating, for PMFs with an amplitude of  $\sqrt{\langle  B^2 \rangle}$=0.4 nG for different spectral indices. The upper panel is $TT$ the lower panel is $EE$. Colors represent the different spectral indices.  }
\label{fig:ClT}
\end{figure*}

\subsection{Combining both effects}
Having discussed the two dissipative effects separately we now analyse the combined effect of PMF heating on the CMB angular power spectra. In Fig.~\ref{fig:ClT}, we again show the $TT$ and $EE$ angular power spectra and their relative difference with respect to the case without PMFs, for fields of $\sqrt{\langle  B^2 \rangle}$=0.4 nG and different spectral indices. We note  how the combination of the two effects results in an impact of both temperature and polarization both on small and large angular scales, with the effect increasing for positive spectral indices. In the next section we will derive the constraints with current CMB data, which are foreground and cosmic-variance limited in temperature, but strongly affected by systematics in polarization. Future CMB polarization dedicated observations will be therefore crucial to fully exploiting the potential of the impact of ambipolar diffusion on the E-mode polarization.

\vspace{-3mm}
\section{CMB constraints on the amplitude of PMFs}
In this section, we derive the constraints with the CMB anisotropy data from {\sc Planck} 2015 release. 
We use the extension of the {\tt CosmoRec} and {\tt Recfast++} codes developed in our previous work  \citep{2015MNRAS.451.2244C} with the regularization of the MHD rate and 
the improved numerical treatment for the ambipolar diffusion discussed in the previous sections.
We  use the \texttt{CosmoMC} \cite{cosmomc} code with the inclusion of the modified recombination codes in order to compute 
the Bayesian probability distribution of cosmological and magnetic parameters.
We vary the baryon density $\omega_{b}=\Omega_{b} h^2$, the cold dark matter density 
$\omega_{c}= \Omega_{c}h^2$ (with $h$ being
$H_0/100\, {\mathrm {km}}\,{\mathrm {s}}^{-1}{\mathrm {Mpc}}^{-1}$), the reionization optical depth 
$\tau$ with a Gaussian prior, the ratio of the sound horizon to the angular diameter distance at decoupling $\theta$, $\ln ( 10^{10} A_S )$, $n_S$ 
and the magnetic parameter $\sqrt{\langle B^2\rangle}$. We either fix $n_{\mathrm B}$ to the values $-2.9 \,, -2 \,, -1 \,, 0 \,, 1 \,, 2$  
or we allow $n_{\mathrm B}$ to vary in the range $[-2.9,2]$. 

Together with cosmological and magnetic parameters we vary the parameters associated to calibration and beam uncertainties,  
astrophysical residuals, which are included in the Planck public likelihood \citep{2016A&A...594A..11P}.
We assume a flat universe, a CMB temperature $T_{\mathrm CMB}=2.725$ K and a pivot scale $k_*=0.05$~Mpc$^{-1}$.
We sample the posterior using the Metropolis-Hastings algorithm \cite{Hastings:1970xy} generating eight 
parallel chains and imposing a conservative Gelman-Rubin convergence
criterion \cite{GelmanRubin} of $R-1 < 0.02$.

We use public {\sc Planck} high-$\ell$ likelihood temperature likelihood \citep{2016A&A...594A..11P} 
combined with the {\sc Planck} lensing likelihood \citep{Ade:2015zua}. 
We use a conservative Gaussian prior for the optical depth $\tau = 0.070 \pm 0.02$ 
in combination with the low-$\ell$ Gibbs Commander likelihood in the range $\ell = [2,29]$ for the low-$\ell$ temperature. 

Note that the likelihood code for the more recent analysis of large angular scales HFI polarisation data \citep{2016A&A...596A.107P,2016A&A...596A.108P} has not been released and we therefore make use only of Planck 2015 data.

\subsection{Constraints with MHD decaying turbulence}

We first present the constraints on the amplitude of PMFs obtained by considering only the heating due to the MHD decaying turbulence term with
the use of the regularized rate.

\begin{figure}
\centering 
\includegraphics[width=0.9\columnwidth]{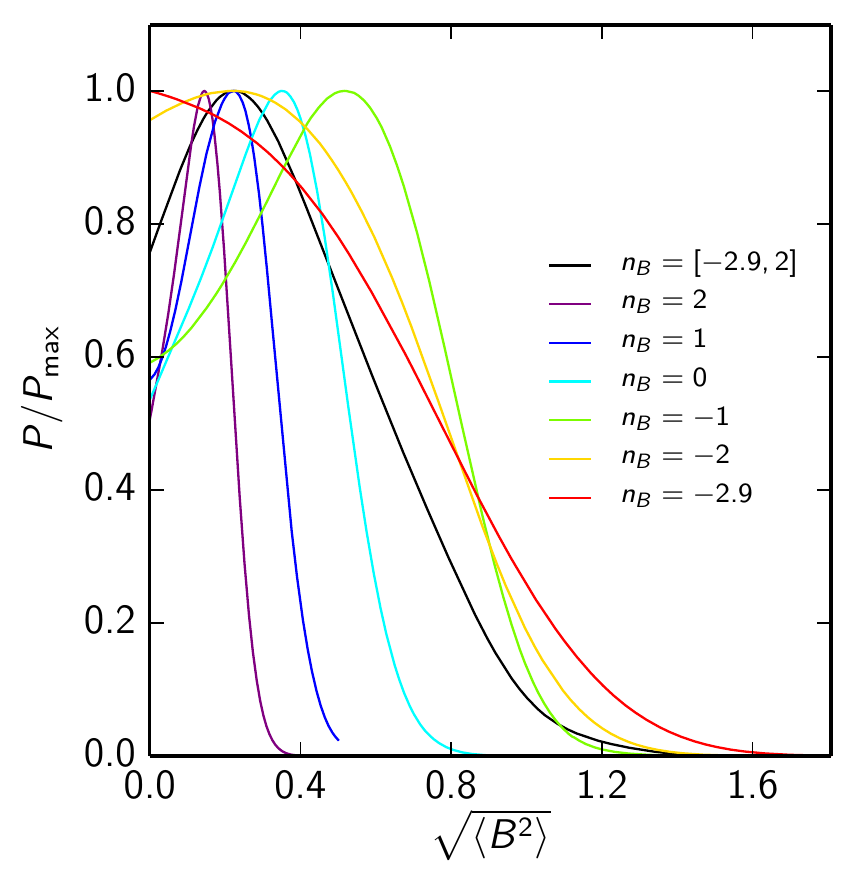}
\caption{One dimensional posterior probabilities considering only MHD for $\langle B^2 \rangle^{1/2}$ for fixed values 
of the spectral index $n_\mathrm{B}$  compared with the case marginalized on $n_\mathrm{B}$ allowed to vary 
in the range $[-2.9,2]$.
}
\label{fig:Results2}
\end{figure}

In Fig. \ref{fig:Results2} we plot the one-dimensional marginalized posterior probabilities 
for $\langle B^2 \rangle^{1/2}$ at different fixed values of the spectral index $n_{\mathrm B}$. 
We also plot the same quantity obtained when $n_{\mathrm B}$ is allowed to vary. 
In the first column of Table \ref{tab_PSP5} we report the 95 \% CL constraints on $\langle B^2 \rangle^{1/2}$ for all the cases considered.
The constraints are at the nano-Gauss level with tighter constraints for positive spectral indices (reduced $\simeq 3-4$ times for $n_{\mathrm B}\simeq 2$ with respect to the quasi-scale invariant case).

\begin{table}
\center
\begin{tabular}{|c|c|c|c|c|c|c|c|c|c|}
\hline
$n_{\rm B}$ & &$\sqrt{\langle  B^2 \rangle} \, (\mathrm{nG}) $   &    \\
\hline
& MHD turbulence & Ambipolar diffusion & Combination\\
\hline
2 & $<0.25$ &  $<0.06$ & $<0.06$\\
\hline
 1 & $<0.37$ &  $<0.12$ &  $<0.13$\\
\hline
0 & $<0.58$ &  $<0.26$ &  $<0.30$ \\
\hline
-1 & $<0.90$ &  $<0.63$ &  $<0.74$\\
\hline
-2 & $<0.93$ &  $<1.88$ &  $<0.90$\\
\hline
-2.9 & $<1.04$ &  $<7.29$ &  $<1.06$\\
\hline
[-2.9,2] & $<0.87$ &  $<2.52$ &  $<0.83$\\
\hline
\end{tabular}
\caption{\label{tab_PSP5}
Comparison of the constraints from the separate effects and their combination.}
\end{table}

\subsection{Constraints with the ambipolar diffusion}

\begin{figure}
\centering
\includegraphics[width=0.9\columnwidth]{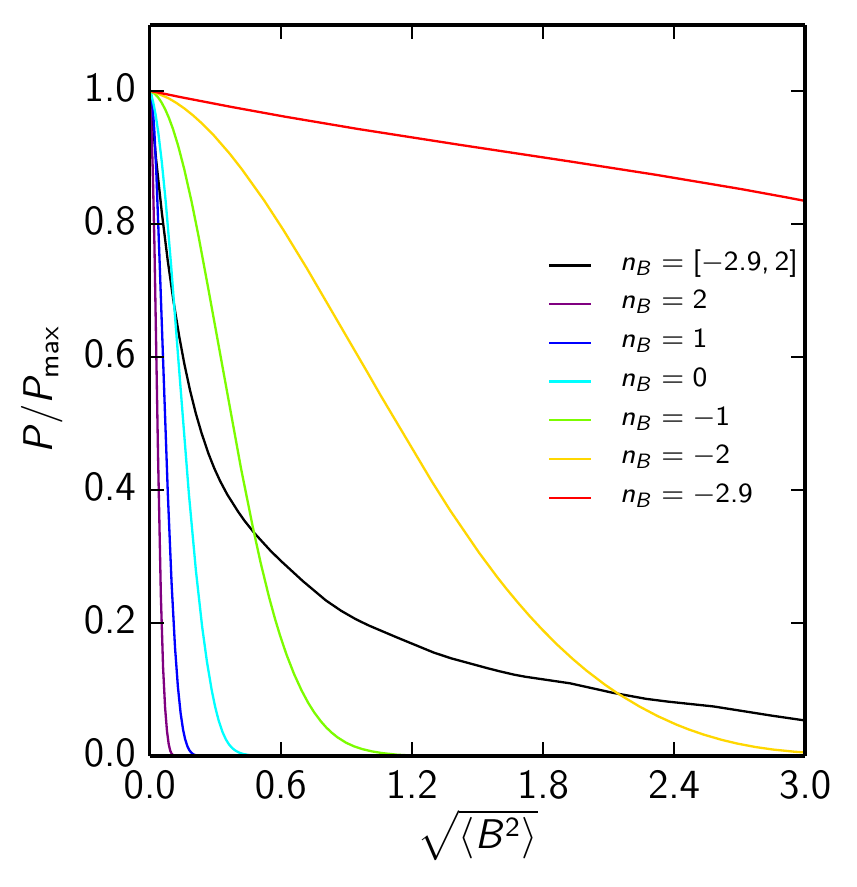}
\caption{One dimensional posterior probabilities considering only ambipolar diffusion for $\langle B^2 \rangle^{1/2}$ for fixed values 
of the spectral index $n_\mathrm{B}$  compared with its corresponding value marginalized on $n_\mathrm{B}$ allowed to vary 
in the range $[-2.9,2]$.
}
\label{fig:Results3}
\end{figure}

In this subsection we presents the constraints on the amplitude of PMFs considering 
only the heating due to the ambipolar diffusion.
In Fig. \ref{fig:Results3} we plot the one-dimensional marginalized posterior probabilities 
for $\langle B^2 \rangle^{1/2}$ at different fixed values of the spectral index $n_{\mathrm B}$. 
We also plot the same quantity obtained when $n_{\mathrm B}$ is allowed to vary. 
In the second column of  Table \ref{tab_PSP5} we report the 95 \% CL constraints on $\langle B^2 \rangle^{1/2}$ for all the cases considered. We note how the ambipolar diffusion gives stronger constraints for growing spectral indices as it is expected from its impact on the CMB angular power spectra. The improvement of the constraint for $n_{\mathrm B}\simeq 2$ with respect to the quasi-scale invariant case is dramatic, reaching a factor $\simeq 100$. This implies that a combination of turbulent MHD and ambipolar diffusion heating is expected to improve the constraints in particular for very blue spectra, as we will see below.

\begin{figure}
\centering
\includegraphics[width=0.9\columnwidth]{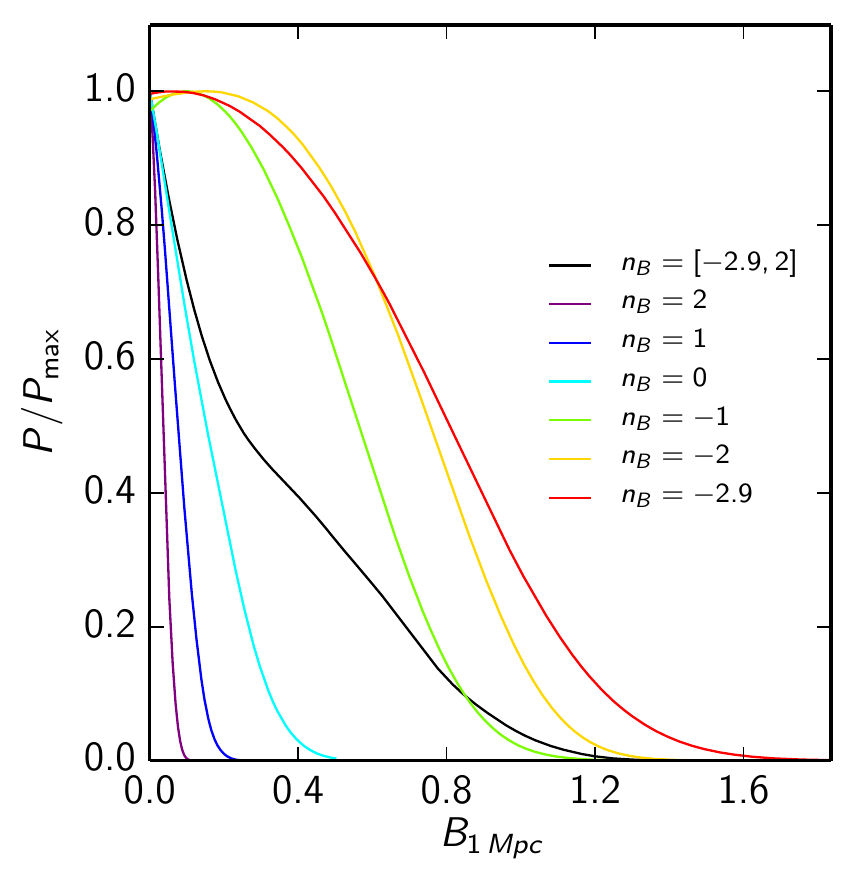}
\caption{One dimensional posterior probabilities for $\langle B^2 \rangle^{1/2}$ considering both the heating effects for fixed values 
of the spectral index $n_\mathrm{B}$  and compared with its corresponding value marginalized on $n_\mathrm{B}$ allowed to vary 
in the range $[-2.9,2]$ .
}
\label{fig:Results4}
\end{figure}

\begin{figure}
\centering
\includegraphics[width=0.9\columnwidth]{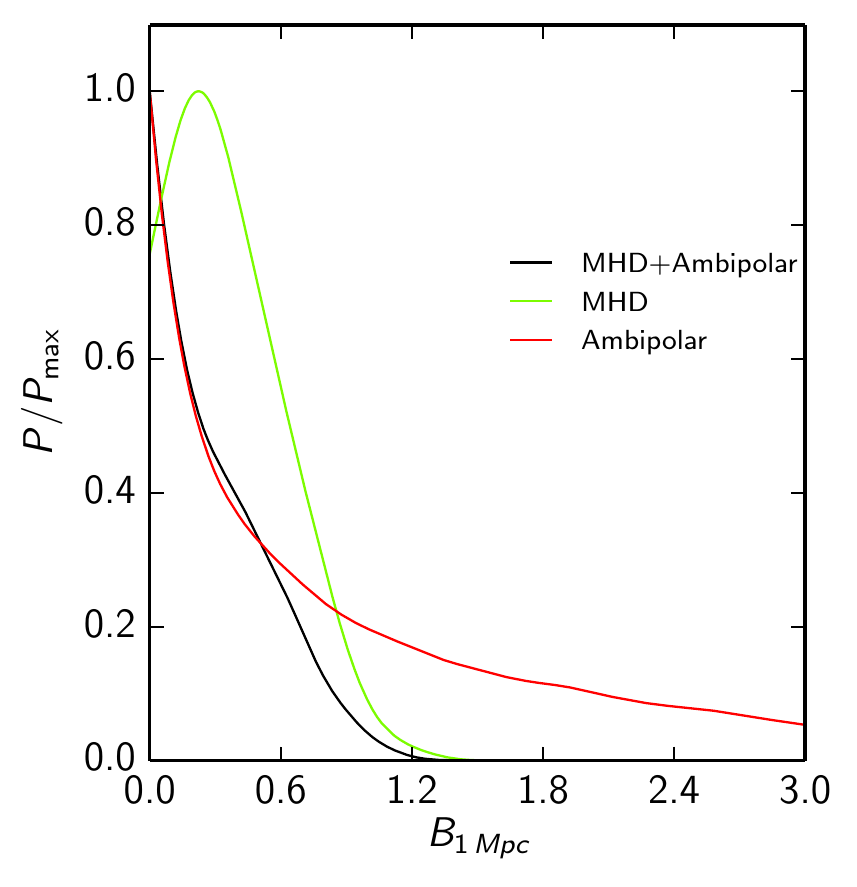}
\caption{Comparison of the constraints marginalized over the spectral index for the three heating cases.
}
\label{fig:Results5}
\end{figure}

\subsection{Constraints including both heating terms}

In this subsection we presents the constraints on the amplitude of PMFs considering  both the effects of the ambipolar diffusion and MHD decaying turbulence.
In Fig. \ref{fig:Results4} we plot the one-dimensional marginalized posterior probabilities 
for $\langle B^2 \rangle^{1/2}$ at different fixed values of the spectral index $n_{\mathrm B}$. 
We also plot the same quantity obtained when $n_{\mathrm B}$ is allowed to vary. 
In the third column of Table \ref{tab_PSP5} we report the 95 \% CL constraints on $\langle B^2 \rangle^{1/2}$ for all the cases considered.
For $n_{\rm B}\lesssim -1$, MHD turbulent heating drives the constraint, while for $n_{\rm B}\gtrsim -1$, ambipolar diffusion become most relevant.

In Fig. \ref{fig:Results5} we present the comparison of the amplitude constraints marginalized over the spectral index. We note how the MHD turbulence has a much sharper posterior distribution compared with the long tail at high amplitudes of the ambipolar diffusion. This effect is mainly due to the strong dependence of the constraints of the ambipolar diffusion with the spectral index. While the MHD turbulence has similar constraining power for all the indices, the ambipolar diffusion is weaker for negative ones resulting in a longer tail.
The combination of the two gives a sharp constraint as shown in Fig. \ref{fig:Results5}, the lower amplitude part of the distribution is dominated by the ambipolar diffusion whereas the higher amplitude side is dominated by the MHD decaying turbulence.

\begin{figure}
\centering
\includegraphics[width=0.98\columnwidth]{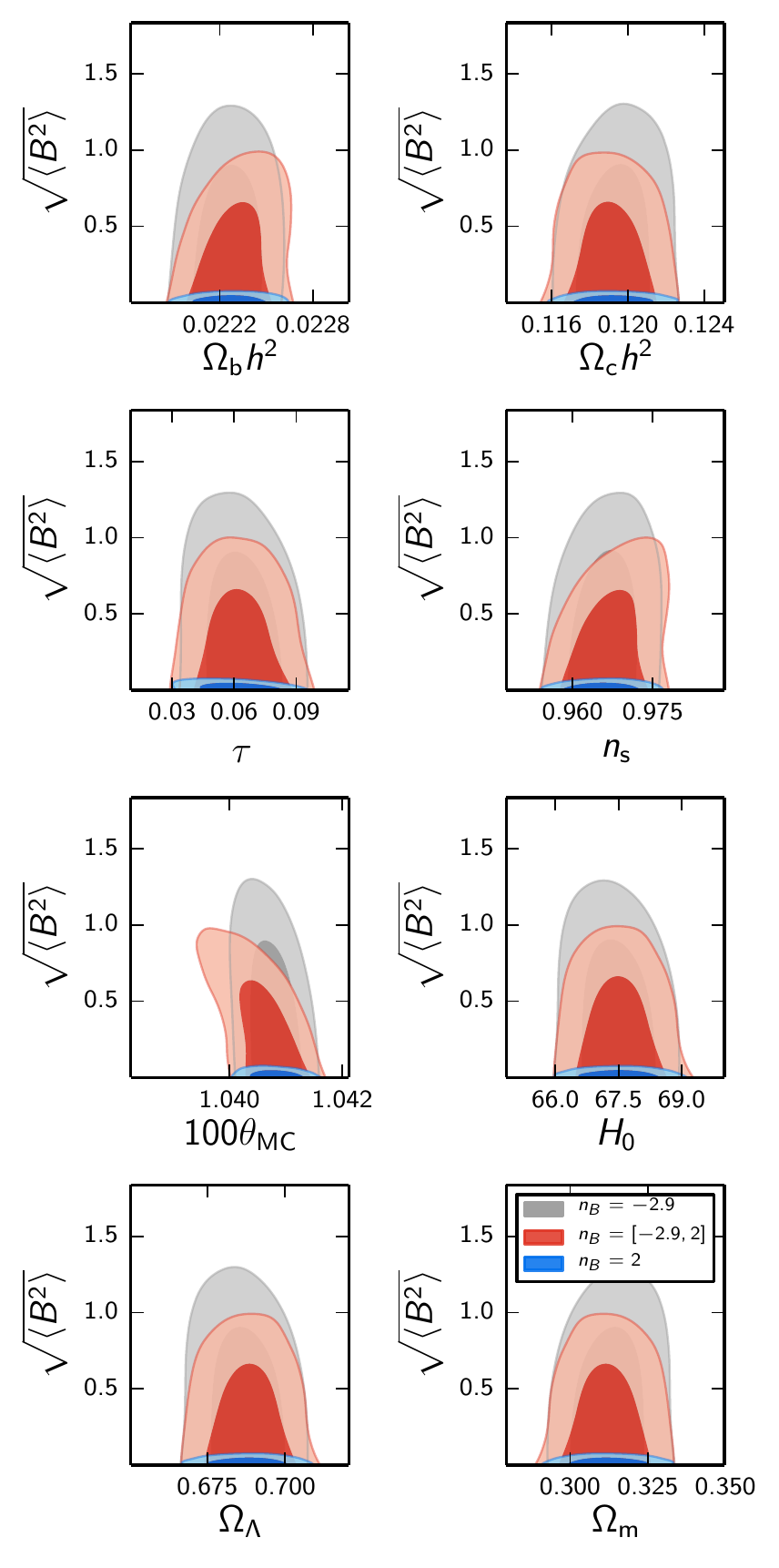}
\caption{Two dimensional posteriors for the amplitude of the fields with the other cosmological parameters. The results are shown for three spectral indices, in blue $n_{\rm B}=2$, in red varying $n_{\rm B}$ in grey is the almost scale invariant $n_{\rm B}=-2.9$.}
\label{2D}
\end{figure}

Finally, in Fig.~\ref{2D} we present the two dimensional posteriors of the amplitude of PMFs with the other cosmological parameters. 
We note the presence of a slight degeneracy with the angular diameter distance $\theta$ especially for the varying spectral index case, this is expected considering the effect of the heating on the recombination.

\section{Discussions}

\begin{figure}
\centering
\includegraphics[width=0.96\columnwidth]{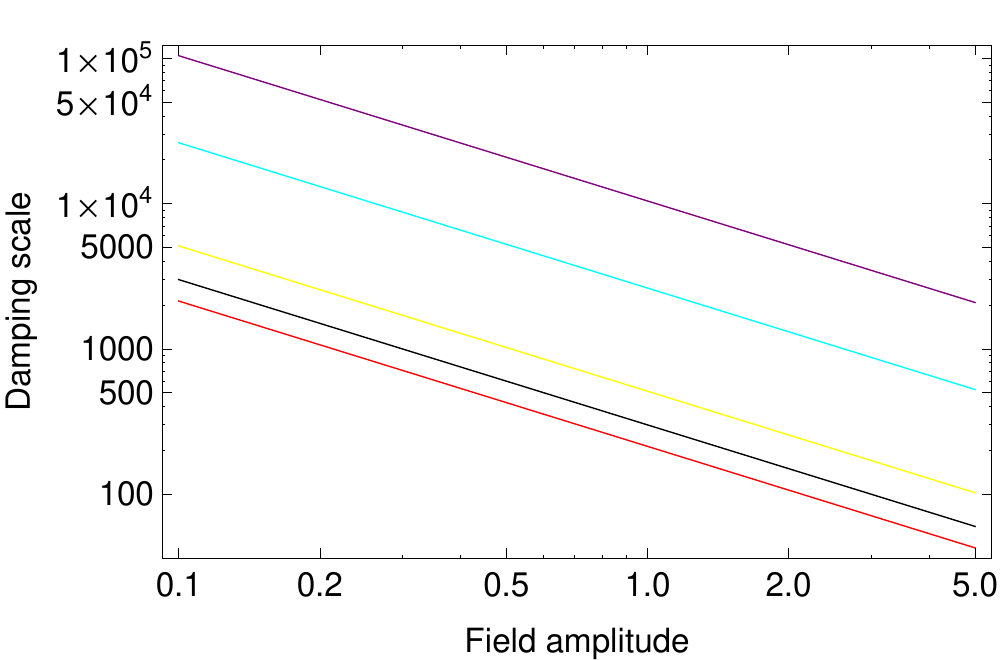}
\caption{Damping scales as function of the integrated PMF amplitude. The black line represents the damping scale in 
Eq.~\eqref{KDKK} vs $B_0 ({\rm nG})$. The other lines represent the damping scale as given by Eq.~\eqref{KD} vs $\sqrt{\langle B^2 \rangle}$ for different values of 
the spectral index: $n_{\rm B}=2$ in purple, $0$ in cyan, $-2$ in yellow and $-2.9$ in red.}
\label{DP}
\end{figure}

We now discuss the dependence of the results presented in Table \ref{tab_PSP5} on the physics at the damping scale. This is tricky and several approaches have been considered in the past.
There is indeed a dependence of both the MHD decaying 
turbulence and ambipolar rates on $k_{\rm D}$ and a dependence on the damping profile in the Lorentz force (compare Eq. (7) with Eqs. (A3-A4) of Appendix A of  \citet{2015MNRAS.451.2244C}). 
We therefore compare the results of Table \ref{tab_PSP5} with the ones obtained by adopting an exponential damping profile as in \citet{2015MNRAS.451.2244C} and \citet{2015JCAP...06..027K}, 
with the following damping scale: 
\begin{equation}
{\bar k}_{\rm D} =\frac{299.66}{(B_0/1\,{\mathrm{nG}})} {\bf \Mpc^{-1}}\,,
\label{KDKK}
\end{equation}
where $B_0$ denotes the integrated amplitude of the stochastic background of PMFs for this second approach to the damping.
Note that ${\bar k}_{\rm D}$ does not depend on the spectral index as the one in Eq.~\eqref{KD} adopted in the previous discussion and has been also used in our previous work \cite{2015MNRAS.451.2244C} for the nearly scale-invariant case. See Fig.~\ref{DP} for a difference between these two damping scales.
We mention that in recent numerical simulations \citep{Trivedi2018} a significantly larger damping scale (smaller $k_{\rm D}$) is found, but leave a more detailed discussion to future work.

We have repeated the previous analysis for this alternative model of damping. 
The qualitative aspects remain similar to the case discussed in Section 2: the MHD term is relevant 
for negative spectral indices, whereas the ambipolar term is for positive ones. Note however that whereas the MHD term leads to constraints 
similar in the two approaches because of the mild dependence on $k_{\rm D}$ of the rate in Eq. (\ref{Eqn:rate}), 
the ambipolar term leads to much looser constraints when this alternative modelling of the damping scale 
is adopted. The constraint with the ambipolar term are indeed of the same order of magnitude of the ones obtained with the MHD term by using this alternative damping envelope.
In Table \ref{tab_KK} we show the results when both the MHD and ambipolar terms are considered: for all values of $n_\mathrm{B}$, the combined constraints are at the nG level. 

Our analysis improves in several ways on \citet{2015JCAP...06..027K}: i) the methodology as described in Section 2, ii) the range of considered PMF spectral indices, which in \citet{2015JCAP...06..027K} was limited to $n_{\mathrm{B}}$ =-2.9, -2.5,-1.5, iii) and the data combination: here we consider the most recent Planck 2015 data, whereas  \citep{2015JCAP...06..027K} used Planck 2013 data. The numerical stability we have achieved removes the large scale instability which could have biased the results especially concerning
the indices with a stronger heating. With these new settings, in contrast to  \citet{2015JCAP...06..027K}, there is almost no variation with the spectral index of the constraints and therefore we do not find tighter bounds for $n_{\mathrm{B}}> -2.9$ as \citep{2015JCAP...06..027K} do and our 95 $\%$CL
constraint $B_0 < 1.1$~nG for $n_{\mathrm{B}}=-2.9$ is more conservative than their corresponding bound: $B_0 < 0.63$~nG. Note that for positive spectral indices 
the constraints from this alternative model of damping are relaxed by a factor 5-20 with respect to the model described in Section 2. The reason for different results in the two approaches is due to the ambipolar term. As already said, the differences could be traced to the different Lorentz force obtained by a different damping envelope or a different damping scale. In order to understand what is the most relevant difference, we have substituted the damping scale in Eq. (\ref{KDKK}) in the sharp-cut off profile for the damping discussed in Section 2 for $n_{\rm B}=2$. 
We obtain $\sqrt{\langle  B^2 \rangle}<1.0 \, \mathrm{nG}$ at 95 \%CL for the combined case, a very similar result to Table  \ref{tab_KK}. 
This means that the most relevant difference is due to choice of $k_{\rm D}$ for the two models of damping discussed here.

It is now interesting to assess the the implications of the
constraints derived in this paper on the amplitude of the stochastic background of PMF smoothed at 1 Mpc, which is commonly adopted in the literature.
Since the damping scale enters in the magnetic field amplitude smoothed $B_\lambda$ as function of the integrated amplitude (see Appendix A), 
$B_\lambda$ can be different for the two dissipation scales in Eq. (\ref{KD}) \citep{Jedamzik1998, Subramanian1998} and in Eq. (\ref{KDKK}) \citep{2015MNRAS.451.2244C,2015JCAP...06..027K}, 
in particular for positive spectral indices, even with equal integrated amplitudes. 
Table \ref{tab_PSPL} shows that for $n_{\rm B} = -2.9$ the constraints on $B_{\mathrm{1 Mpc}}$ from the two different damping envelopes are similar and 
of the same of order of magnitude of the constraints on the integrated amplitude. This can be understood by realizing that for quasi-scale independent power spectrum the increase of $\langle  B^2 \rangle$ (which simply is a proxy for the total PMF energy density) caused by small scales is logarithmic, and hence $B_{\mathrm{1 Mpc}}\simeq \sqrt{\langle  B^2 \rangle}$.

For $n_{\rm B}=2$ instead, the energy density is dominated by modes around the damping scales. In this case, we see from Table~\ref{tab_PSPL} that the constraint on $B_\lambda$ with the damping scale in Eq.~\eqref{KD} is tighter than the one obtained with the alternative damping by several orders of magnitude. To a large extend this is due to the large disparity of the damping scale ($\lambda_{\rm D}\simeq 1- 10\,{\rm kpc}$) and the smoothing scale ($\lambda =1\,{\rm Mpc}$), as can be seen from Eq.~\eqref{app:A3}. In the most conservative case, the window for PMF between the CMB bounds and the lower limit due to the interpretation of non-observation of GeV gamma-ray emission in intergalactic medium is severely squeezed for $n_{\rm B}=2$. 
The tightest constraint obtained with Eq.~\eqref{KD} would instead completely rule out the causal case $n_{\rm B}=2$ in combination with the lower limit derived from high-energy observations in the intergalatic medium.

In this paper we limited our analysis to non-helical magnetic fields. Helical magnetic fields may have a different effect on the ionization history with respect to non-helical ones. Helicity may affect the ambipolar diffusion through the contribution of the helical symmetric part of the Lorentz force \citep{Ballardini:2014jta} and it may affect the evolution of the MHD turbulence \citep{Wagstaff:2015jaa} and modify the time evolution of the magnetic energy density \citep{Saveliev:2013uva}. We leave the treatment of helical magnetic fields to future work.

\begin{table}
\center
\begin{tabular}{|c|c|c|c|c|c|c|c|c|c|}
\hline
$n_{\rm B}$ & 2 & -2.9 & [-2.9,2]\\
\hline
$B_0 \, (\mathrm{nG}) \,[{\bar k}_{\rm D}] $ & $<0.95$ & $<1.10$& $<0.91$\\
\hline
\end{tabular}
\caption{\label{tab_KK}
Constraints from the combined effects for the alternative model of the damping profile, $B_0$.}
\end{table}

\section{Conclusions}

We have obtained the constraints on the integrated amplitude of PMFs due to their 
dissipation around and after recombination caused by the MHD decaying turbulence and the ambipolar diffusion.
We have improved our previous treatment by including a regularization of the heating rate due to the MHD decaying turbulence 
which is particularly important for stochastic background of PMFs with a positive spectral index.
At the same time, we have also improved the numerical treatment of the ambipolar diffusion allowing for the stability of the
numerical code, again for stochastic background of PMFs with positive spectral indices.
These improvements have allowed to constrain the integrated amplitude of PMFs for different spectral indices, extending our previous studies restricted 
to the nearly scale-invariant case \citep{2015JCAP...06..027K, Ade:2015cva,2015MNRAS.451.2244C}. 

The results of the three analysis which considered separately the heating by MHD decaying turbulence and ambipolar diffusion and their combination are summarized in Table \ref{tab_PSP5} for a regularization of the integrated amplitude by a sharp cut-off. Our results show that both MHD decaying turbulent and ambipolar effects need to be taken into account, the first one being important for negative spectral index and the second for positive spectral index. For a sharp cut-off the combined constraint from MHD and ambipolar is of the order of nG for the scale-invariant case as in \citep{Ade:2015cva}, and becomes tighter with a larger spectral index reaching 
$\sqrt{\langle B^2 \rangle} < 0.06$~nG (95 \% CL) for $n_{\rm B}=2$. These constraints on PMFs from the ionization history are the tightest ones for any single spectral index. Thanks to our numerical improvements we have also been able to derive the constraints on the integrated amplitude when the spectral index is allowed to vary, obtaining $\sqrt{\langle  B^2 \rangle} <0.83 $ nG (95\% CL) [see Fig.~\ref{fig:Results5}]. 

We have also investigated how the PMFs heating effects are sensitive to the physics at the damping scale. We have shown how two proposed damping scales, Eq.~\eqref{KD} and Eq.~\eqref{KDKK}, usually adopted in the literature, lead to a different magnitude of the effect induced by the ambipolar term on the CMB anisotropy power spectra, in particular for positive spectral indices. As a consequence, the constraints obtained on the integrated amplitude of PMFs, and even more on the smoothed amplitude on 1 Mpc, depend on 
the physics at the damping scale, which deserve further investigation. In the future, some of these aspects can be clarified with detailed numerical MHD simulation that track the evolution of the PMF across the recombination era \citep{Trivedi2018}.

We also note that although recently refined computations of the magnetic heating rates due to MHD turbulence have become available \citep{Trivedi2018}, here we improved the treatment remaining within the framework first introduced by \citet{Sethi2005}. However, the improved heating rate computations show a direct dependence of the onset of heating on the magnetic field amplitude and spectral index. We anticipate this to affect the overall constraints, but a more detailed study is left to future work. 

Our results show that the effect of PMFs on the ionization history provides stronger constraints than purely gravitational effects under the same assumptions of ideal MHD and a damping scale comoving in time. The impact on the E-mode polarization makes this effect a target for current and future CMB experiments which are expected to provide a nearly cosmic variance limited E-mode measurement. The constraints by the gravitational effect are expected to improve thanks to the separation of the primary signal from secondary anisotropies/foreground residuals  at very high multipoles in temperature and on the future B-mode measurements. The ionization history and gravitational effects caused by PMFs therefore have different and complementary capabilities and prospects. 


\small
\section*{Acknowledgments}
We thank Luke Hart for useful discussions about numerical issues. 
DP and FF acknowledge support by the "ASI/INAF Agreement 2014-024-R.0 for the Planck LFI Activity of Phase E2 and the financial support by ASI Grant 2016-24-H.0. JC is supported by the Royal Society as a Royal Society University Research Fellow at the University of Manchester, UK. JARM acknowledges financial support from the Spanish Ministry 
of Economy and Competitiveness (MINECO) under the projects AYA2014-60438-P and AYA2017-84185-P.

{
\vspace{-0mm}
\small
\bibliographystyle{mn2e}
\bibliography{Lit}
}

\begin{appendix}

\section{Constraints on smoothed magnetic field amplitude}

In most of the literature, constraints on a stochastic background of PMFs 
are reported on the amplitude smoothed at 1 Mpc scale, which is a quantity closer to astrophysical 
observations of large scale magnetic fields. It is therefore interesting to understand our results for the integrated amplitude 
in terms of the smoothed amplitude $B_{\lambda}$, which is defined as:
\begin{equation}
B^2_\lambda = \int_0^{\infty} \frac{d{k \, k^2}}{2 \pi^2} e^{-k^2 \lambda^2} P_B (k).
\end{equation}
The smoothed amplitude $B_\lambda$ is related to the integrated amplitude by 
\begin{equation}
\langle B^2\rangle = B^2_\lambda \frac{2\,k_{\rm D}^{n_{\rm B}+3}\lambda^{n_{\rm B}+3}}{(n_{\rm B}+3)\Gamma\Big(\frac{n_{\rm B}+3}{2}\Big)}\,,
\end{equation}
for the first damping envelope and by
\begin{equation}
\label{app:A3}
B^2_\lambda = B_0^2 \, 2^{(n_{\rm B}+3)/2}/(k_{\rm D} \lambda)^{n_{\rm B}+3}
\end{equation}
for the second damping envelope.

In Table \ref{tab_PSPL} we report the implications for $B_\lambda$ from the our results on the integrated amplitude. 
A cautionary note must be considered when discussing these results.
The derived constraints on the smoothed amplitude seems very sensitive to the model of damping, in particular for positive spectral index.
Nevertheless, the resulting constraints are extremely tight for positive $n_{\rm B}$ compared to those obtained with the 
gravitational contribution only. As a comparison, we remind that the 95~\% CL Planck 2015 upper bound on the smoothed amplitude is $B_\lambda < 0.011$~nG for $n_{\rm B}=2$ derived from gravitational effects \citep{Ade:2015cva}.

\begin{table}
\center
\begin{tabular}{|c|c|c|c|c|c|c|c|c|c|}
\hline
$n_{\rm B}$& $B_{1 \mathrm{Mpc}} \, (\mathrm{nG}) $\ & $B_{1 \mathrm{Mpc}} \, (\mathrm{nG})\, [{\bar k}_{\rm D}]$\\
\hline
2 &   $<5.22\times 10^{-16}$&  $<1.13\times 10^{-6}$\\
\hline
-2.9 & $<0.76$ &  $<0.84$\\
\hline
\end{tabular}
\caption{\label{tab_PSPL}
Constraints from the combined effect for different spectral indices  with the $B_{1\, {\rm Mpc}}$ parametrization.}
\end{table}
\end{appendix}

\end{document}